\documentclass[pdftex,onecolumn,epjc3]{svjour3} 
\pdfoutput=1
\usepackage{graphicx,subcaption}  
\usepackage{dcolumn}   
\usepackage{amsmath, amssymb}        
\usepackage{color}
\usepackage{url,hyperref,cite}
\usepackage{lineno}
\usepackage{siunitx}
\usepackage{booktabs}
\usepackage{multirow}
\usepackage{enumitem}
\usepackage[usenames,dvipsnames]{xcolor}
\usepackage{SymbolsNLOPrecision}
\usepackage{SHFeynGraphUtils}
\usepackage{soul} 
\usepackage{mathtools, cuted} 
\usepackage{xcolor}
\usepackage[absolute]{textpos} 
\usepackage{url}

\def\ifb{\rm~fb^{-1}}
\def\iab{\rm~ab^{-1}}

\newcommand{\invab}{{\rm ~ab^{-1}}}
\newcommand{\GeV}{{\rm~GeV}}
\newcommand{\TeV}{{\rm~TeV}}
\newcommand{\mgamc}{\texttt{MG5\_aMC@NLO}}
\newcommand{\mgpy}{\mgamc\texttt{+PY8}}
\newcommand{\sherpa}{\texttt{Sherpa}}
\newcommand{\hathor}{\texttt{HATHOR}}
\newcommand{\confirm}[1]{{\color{black} #1}}

\usepackage{geometry}
\hypersetup{
   colorlinks=true,       
   linkcolor=blue,        
   citecolor=red,         
   filecolor=magenta,     
  }

\journalname{Eur. Phys. J. C}

\begin{document}
\title{Monojet Signatures from Heavy Colored Particles: Future Collider Sensitivities and Theoretical Uncertainties}

\begin{textblock}{13.625}(1,0.75) 
\hfill KEK-TH-2050, IPPP/18/31, SLAC-PUB-17265
\end{textblock}

\subtitle{}

\titlerunning{Monojet Signatures from Heavy Colored Particles}        

\author{Amit Chakraborty\thanksref{e1,addr1}\and
	Silvan Kuttimalai\thanksref{e2,addr2}\and
	Sung Hak Lim\thanksref{e3,addr1}\and
	Mihoko M. Nojiri\thanksref{e4,addr1,addr3,addr4}\and
	Richard Ruiz\thanksref{e5,addr5}
}

\thankstext{e1}{e-mail: amit@post.kek.jp}
\thankstext{e2}{e-mail: silvan@slac.stanford.edu}
\thankstext{e3}{e-mail: sunghak.lim@kek.jp}
\thankstext{e4}{e-mail: nojiri@post.kek.jp}
\thankstext{e5}{e-mail: richard.ruiz@durham.ac.uk}

\institute{
\label{addr1} Theory Center, IPNS, KEK, Tsukuba, Ibaraki 305-0801, Japan \and
\label{addr2} SLAC National Accelerator Laboratory, 2575 Sand Hill Road, Menlo Park, CA 94025, USA \and
\label{addr3} The Graduate University of Advanced Studies (Sokendai), Tsukuba, Ibaraki 305-0801, Japan \and
\label{addr4} Kavli IPMU (WPI), University of Tokyo, Kashiwa, Chiba 277-8583, Japan  \and
\label{addr5} Institute for Particle Physics Phenomenology {\rm(IPPP)}, Department of Physics, Durham University, Durham, DH1 3LE, UK 
}

\date{Received: date / Accepted: date}
%
%
\maketitle
\begin{abstract}  
In models with colored particle $\mathcal{Q}$ that can decay into a dark matter candidate $X$,
the relevant collider process $pp\to \mathcal{Q}\bar{\mathcal{Q}}\to X\bar{X}+$jets
gives rise to events with significant transverse momentum imbalance.
When the masses of $\mathcal{Q}$ and $X$ are very close, the relevant signature becomes monojet-like,
and Large Hadron Collider (LHC) search limits become much less constraining. 
In this paper, we study the current and anticipated experimental sensitivity to such particles at the
High-Luminosity LHC at $\sqrt{s}=14$ TeV with $\mathcal{L}=3$ ab$^{-1}$ of data and the proposed
High-Energy LHC at $\sqrt{s}=27$ TeV  with $\mathcal{L}=15$ ab$^{-1}$ of data. 
We estimate the reach for various Lorentz and QCD color representations of $\mathcal{Q}$. 
Identifying the nature of $\mathcal{Q}$ is very important to understanding the physics behind the monojet signature. 
Therefore, we also study the dependence of the observables built from the $pp\to\mathcal{Q}\bar{\mathcal{Q}} + j $ process
on $\mathcal{Q}$ itself.
Using the state-of-the-art Monte Carlo suites \texttt{MadGraph5\_aMC@NLO+Pythia8} and \texttt{Sherpa}, 
we find that when these observables are calculated at NLO in QCD with parton shower matching and multijet merging, 
the residual theoretical uncertainties are comparable to differences observed when varying the quantum numbers of $\mathcal{Q}$ itself.
We find, however, that the precision achievable with NNLO calculations, where available, can resolve this dilemma.
\keywords{Standard Model \and BSM physics \and LHC \and Jets \and QCD}
\end{abstract}
%
\section{Introduction}
The nature of dark matter (DM) remains one of the outstanding mysteries in the particle physics today.
Among the many possible particle candidates known~\cite{Feng:2010gw},  weakly 
interacting massive particles (WIMP) are arguably the most theoretically motivated and well-studied 
scenarios; for a recent review, see \cite{Abercrombie:2015wmb}.
WIMPs with masses in the range of 10 GeV - 20 TeV~\cite{Plehn:2017fdg} can be stable on the age of the universe, 
and once they are in thermal equilibrium in the early universe remain so, even after decoupling occurs. 
Moreover, such stable particles are naturally present in many beyond the Standard Model (BSM) frameworks.
For example:
In weak-scale supersymmetric (SUSY) theories~\cite{Nilles:1983ge,Haber:1984rc}, if one assumes R-parity conservation, 
then then lightest SUSY particle is stable, and hence is a potential component of DM.
In the universal extra dimension (UED) models \cite{Appelquist:2000nn}, 
the lightest Kaluza-Klein  excitations of neutral electroweak bosons can be viable DM candidates.

Notably, a number of these new physics scenarios also involve additional heavy colored particles, $\newResonance$, that couple to DM candidate(s).
Hence, if $\newResonance$ and DM are kinematically accessible at the Large Hadron Collider (LHC), or 
its potential successors, such as the High Energy (HE)-LHC, then it may be possible to study DM in a laboratory setting.
In particular, once produced and if allowed, $\newResonance$ can decay into DM and SM particles leading to a plethora of interesting signatures at the LHC.

At hadron colliders, search strategies for these hypothetical colored 
particles usually involve investigating jets and leptons produced in association 
with final-state DM candidates manifesting as large missing transverse energy ($\met$).
In the context of simplified SUSY models, such signatures are now strongly constrained by LHC data if $\newResonance$ 
has a mass around 1~TeV and below~\cite{Sirunyan:2017mrs,Sirunyan:2017xse,Aaboud:2017aeu,Aaboud:2017bac}.
Such constraints on $\newResonance$, however,  can be circumvented.
One of the most celebrated examples of this is the compressed spectrum scenario~\cite{Martin:2007gf,Fan:2011yu,Murayama:2012jh}.
In this situation, the DM and new colored particles have a small mass splitting.
Consequently, the visible decay products in the  $\newResonance\to$DM+SM process do not have 
sufficient momenta to be readily distinguished from SM backgrounds.
In other words, the compression of the mass spectrum constrains the visible decay products of  $\newResonance$
to possess such low momenta that they fail experimental selection criteria.
This leads to significantly smaller selection and acceptance efficiencies, and hence significantly weaker bounds on heavy particle masses.

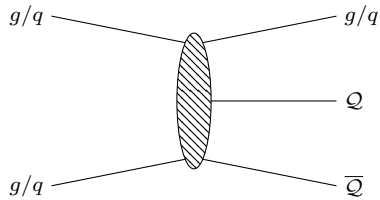
\begin{figure}
\begin{center}
\begin{tikzpicture}[baseline={([yshift=-.5ex]current bounding box.center)},vertex/.style={anchor=base,
    circle,fill=black!25,minimum size=18pt,inner sep=2pt},scale=0.75]
//Interacting parton
\draw (0.5, 1.5) -- (3.0, 1.0);
\draw (0.5, -1.5) -- (3.0, -1.0);
// Hard Radiation
\draw (3.0, 0) -- ( 5.5 , 0.0 );
//Hard process
\draw (3.0,1.0) -- (5.5,1.5);
\draw (3.0,-1.0) -- (5.5,-1.5);
// p p > g j intermediate vertex
\draw [draw=none, fill=white] (3.0, 0) ellipse (0.3 and 1.2);
\draw [pattern=north west lines] (3.0, 0) ellipse (0.3 and 1.2);
//Initial state
\node[draw=none, anchor=east] at (0.5,1.5) {$g/q$};
\node[draw=none, anchor=east] at (0.5,-1.5) {$g/q$};
// intermediate
\node[draw=none, anchor=west] at (5.5,1.5) {$g/q$};
\node[draw=none, anchor=west] at (5.5,-1.5) {$\newResonanceBar$};
\node[draw=none, anchor=west] at (5.5,0) {$\newResonance$};
\node[draw=none, anchor=west] at (4.5,2.5) {\hphantom{$g/q$}};
\end{tikzpicture}
\end{center}
\caption{
Diagrammatic depiction of $\QQbar$ pair production with an extra hard QCD radiation in $pp$ collisions. 
}\label{fig:GraphBSM}
\end{figure}

Even though a compressed mass spectrum represents a special corner of a typical BSM parameter space, 
the most attractive feature of this situation is that it allows relatively light, 
colored DM partners in light of present-day LHC data. This is particularly true for 
stops ($\stp$) and gluinos ($\go$) in SUSY~\cite{Schofbeck:2016oeg,Sirunyan:2018iwl,CMS-PAS-SUS-17-005,ATLAS:2016kts}.
If a compressed scenario is realized in nature, then one can experimentally resolve the soft, i.e., low $\pt$, visible decays of $\Q$ 
by recoiling against a relatively hard, i.e., high $\pt$, electroweak or QCD radiation that, in its own right, 
is sufficiently energetic to satisfy trigger criteria.
One such process, shown diagrammatically in \figref{fig:GraphBSM} and the focus of this study, 
is the inclusive monojet plus $\met$ collider signature \footnote{Note that unlike the exclusive monojet signature, the current search strategy for the inclusive signature permits topologies with up to four analysis-quality jets~\cite{atlas-conf-2017-060}.}.

Were evidence for the new particle $\Q$ established at the LHC, or a successor experiment,
it would be crucial to determine $\Q$'s properties, especially its mass, spin, and color representation.
Generically, such a program would involve investigating various collider observables that can discriminate against possible candidates for $\Q$.
For example: cross sections are highly sensitive to the aforementioned spacetime and internal quantum numbers.
Consider the cases of a scalar top $\stp$ (color triplet), a spin-1/2 top partner $\tp$ (color triplet), 
and spin-1/2 gluino $\go$ (color octet).
For a fixed mass, i.e., $m_{\stp} =m_{\tp} =m_{\go}$, the pair production cross sections for these particles exhibit the hierarchy
\begin{equation}
\sigma(\stst) \ll \sigma(\tptp) \ll \sigma(\gogo). 
\end{equation}
Conversely, for a fixed cross section, i.e., $\sigma(\stst)=\sigma(\tptp)=\sigma(\gogo)$, one finds that
\begin{equation}
m_{\go} > m_{\tp} > m_{\stp}.
\end{equation}
This implies, however, that were a monojet cross section measured, 
the result could be replicated by different $\Q$ scenarios by a simple tuning of mass $m_{\Q}$.
In another way, one cannot constrain the mass of $\Q$ from cross section measurements alone without first asserting its color representation and spin.
Quantitatively, this is more nuanced due to fact that leading order (LO) calculations are poor approximations for QCD processes,
even when using sensible scale choices.
Using Ref.~\cite{Alwall:2014hca}, one can easily verify that, like the top quark~\cite{Nason:1987xz,Beenakker:1988bq}, 
QCD corrections at next-to-leading order (NLO) increase the production cross section $\sigma(\tptp)$  by $\mathcal{O}(50)\%$ for TeV-scale $\tp$.
This is the case at both $\sqrt{s} = 14$ and 27 TeV, and 
despite scale uncertainties  at LO and NLO  spanning $\mathcal{O}(20-30)\%$ and $\mathcal{O}(10)\%$, respectively, for $\sigma(\tptp)$.
Moreover, it is well-known that next-to-next-to-leading order (NNLO) corrections are non-negligible for SM top production~\cite{Czakon:2013goa}.
It is also known that such large theoretical uncertainties can greatly limit the interpretation of the experimental results,
particularly in searches for so-called top-philic dark matter~\cite{Arina:2016cqj}.

The situation, however, is more hopeful following the advent of general-purpose, precision Monte Carlo (MC) event generators.
With software suites such as \texttt{Herwig}~\cite{Bellm:2015jjp}, 
\texttt{MadGraph5\_aMC@NLO+Pythia8} (\mgpy) ~\cite{Alwall:2014hca,Sjostrand:2014zea}, 
and \texttt{Sherpa}~\cite{Gleisberg:2008ta}, 
automated event generation at NLO in QCD with parton shower (PS) matching is now possible for both SM and BSM~\cite{Degrande:2014vpa} processes.
Not only can one now readily include potentially important $\mathcal{O}(\alpha_s)$ corrections to cross sections normalizations,
but parton showers augment fixed order predictions with resummed corrections to at least the leading logarithmic (LL) level.
As a consequence, such observables like the associated jet multiplicity in the monojet process,
an exclusive observable that is critical to search strategies, is automatically modeled at LO+LL accuracy. 
This is the lowest order at which the quantity is qualitatively correct.
In light of the availability of such sophisticated technology, 
one is now in position to systematically investigate the impact of QCD corrections on the inclusive monojet process.

In this report, we perform such a dedicated precision study on the inclusive monojet signature in the context of a compressed mass spectrum.
As mentioned, observables associated with this process are highly sensitive to the mass, spin, and color representation of the mediating states.
Hence, we consider benchmark models with representative mass, spin and color configurations for $\Q$, 
with $\Q \in \{\stp,\tp,\go\}$.
Our study is aimed at the HL-LHC, assuming $\mathcal{L} = 3\invab$ at $\sqrt{s}=14$ TeV, 
and the HE-LHC, assuming $\mathcal{L} = 15\invab$ at $\sqrt{s}=27$ TeV.
We show how the monojet search strategy can be used to identify the nature of $\Q$
from the $\pt$ dependence of the leading jet in $\QQbar$ pair production. 
We also estimate the  precision  required to distinguish these new physics scenarios. 
We quantitatively discuss various sources of theoretical uncertainty, including event generator dependence. 
Although we do not discuss the second jet 
distribution intensively in this paper, azimuthal angle correlation of the first and second jet contains information of the spin of $\Q$, 
as discussed in~\cite{Hagiwara:2013jp,Mukhopadhyay:2014dsa}.

The remainder of this study continues in the following manner:
In \sectionref{sec:numeric} we provide in-dept detail of our computational setup.
In \sectionref{sec:Monojet}, we discuss observed and expected sensitivity of monojet searches at present and hypothetical future facilities,
and address various theoretical uncertainties in \sectionref{sec:theoUncertainty}.
A brief outlook on the impact of this work is discussed in \sectionref{sec:outlook}, and we conclude in \sectionref{sec:summary}.

\section{Computational and Theoretical Setup}\label{sec:numeric}
Systematic studies of QCD radiation in the production of hypothetical, TeV-scale colored particles 
are now possible due to the availability of precision, general-purpose MC event generators.
In practice, this nontrivial task is handled by using several individually published formalisms and software packages 
that have largely been integrated into a single framework or well-specified tool chain sequence. 
In this section, we describe our computational and theoretical setups for modeling $\QQbar$ production 
with various associated jet multiplicities at LO+PS and NLO+PS in \mgpy~and \sherpa.
For numerical results, readers can go directly to \sectionref{monojet}.

The section continues as follows:
In \sectionref{sec:SetupModels}, we briefly summarize the $\Q$ we consider from representative BSM scenarios.
In \sectionref{sec:SetupMatching}, we enumerate the several methods for 
incorporating additional QCD radiation into $\QQbar$ production 
that we employ, and briefly note their main features and formal accuracies.
We describe our setup for \mgpy~and \sherpa, respectively, in \sectionrefs{sec:SetupMGPY}{sec:SetupSherpa},
and our detector simulation in \sectionref{sec:SetupReco}.
In \sectionref{sec:SetupSM}, we summarize the SM inputs.

\subsection{Framework for New Heavy Colored Particles}\label{sec:SetupModels}

In this analysis, we consider three benchmark BSM candidates for $\newResonance$: a stop squark $\stp$, a gluino $\go$, and a fermionic top partner $\tp$.
To model these states in $pp$ collisions at our desired accuracy, 
we use the NLO in QCD-accurate Universal FeynRules Object (UFO)~\cite{Alloul:2013bka,Degrande:2011ua} model libraries available
from the \texttt{FeynRules} model database \cite{ufo_nlo_database}.
The $\mathcal{O}(\alpha_s)$ counterterms required for NLO computations and contained in these libraries are generated 
with \texttt{FeynRules} \cite{Alloul:2013bka}, using \texttt{NLOCT} \cite{Degrande:2014vpa} and \texttt{FeynArts} \cite{Hahn:2000kx}. 
For illustrative purpose, we choose three mass values for $\Q$, namely $m_{\Q}$ = 400 GeV, 600 GeV, and 800 GeV. 
We note that as the spacetime and $\mathrm{SU}(3)_c$ quantum numbers for $\tp$ are identical to those of the SM top quark, 
several publicly available calculations can be adapted in straightforward ways for $\tp$.
This includes total cross section predictions for inclusive $\pptptp$ production at NNLO in QCD, 
which we obtain using the \hathor~package~\cite{Aliev:2010zk}. In Sec \ref{sec:ident}, where we study the dependence of the selected observables 
on the color and spin representations of $\newResonance$, we also briefly consider the 
well-studied~\cite{Kribs:2007ac,Plehn:2008ae,GoncalvesNetto:2012nt} case of a scalar gluon $(\sigma)$ at $m_\sigma = 600\GeV$.

\begin{table*}
\caption{Summary of signal particles, their SU$(3)_c$ and Lorentz representations (Rep.), and decay mode to stable DM candidate $(\dm)$.}
\label{tab:model}
\hphantom{\cite{Fuks:2016ftf,Degrande:2015vaa,Degrande:2014sta,ufo_nlo_database}}
\begin{center}
\begin{tabular}{rccllll}
\hline\noalign{\smallskip}
Particle name & 	& Color Rep.    &   Lorentz Rep. &       Decay  	& UFO Refs. \\ 
\noalign{\smallskip}\hline\noalign{\smallskip}
Fermionic Top partner 	& $(\tp)$  	&  $\mathbf{3}$ &   Dirac fermion	&  $\q + \dm$   & \cite{Fuks:2016ftf,ufo_nlo_database} \\
Top squark  		& $(\stp)$	&  $\mathbf{3}$ &   Complex scalar	&  $t^{*}\dm\to b\qqbar'+\dm$ & \cite{Degrande:2015vaa,ufo_nlo_database} \\
Gluino 			& $(\go)$       &  $\mathbf{8}$ &   Majorana fermion	&  $\qqbar + \dm$ & \cite{Degrande:2015vaa,ufo_nlo_database}   \\
Scalar Gluon 		& $(\sigma)$    &  $\mathbf{8}$ &   Real Scalar		&  \sout{\hphantom{000}} & \cite{Degrande:2014sta,ufo_nlo_database} \\
\noalign{\smallskip}\hline
\end{tabular}
\end{center}
\end{table*} 

    Furthermore, in order to focus on the ISR from $\Q\Qbar$ production and also to use traditional analysis techniques, 
    we assume that the decay of $\Q$ is prompt, with its characteristic lifetime ($\tau_{\Q}$) satisfying $d_0 = \beta c\tau_{\Q} \ll 100~\mu m$, 
    where $\beta$ denotes the relative velocity of $\Q$.
    As a result, $\Q$'s total width must respect the boundary $ \Gamma_{\Q} = \hbar / \tau_{\Q} \gg 2~{\rm meV}$. 
    In realizations of the $\Q$ we consider, e.g., Refs.~\cite{Fuks:2016ftf,Degrande:2015vaa,Degrande:2014sta},
    this stipulation on $\Gamma_{\Q}$ is readily satisfied by large regions of the models' phenomenologically relevant parameter spaces.    
    We also assume that the $\Q\to X+{\rm SM}$ decay is $\Q$'s dominant decay mode. 
    By virtue of DM being weakly coupled, 
    this implies that ${\Q}$'s total width scales as $\Gamma_{\Q}\sim g_{\Q}^2 m_{\Q}$, with an effective $\Q$-DM coupling $g_{\Q}\ll1$.    
    This ensure that the width-to-mass adheres to the inequality
    \begin{equation}
    1 \gg \frac{\Gamma_{\Q}}{m_{\Q}} \gg \left(\frac{2~{\rm meV}}{m_{\Q}}\right) \approx 3\times10^{-15} \times \left(\frac{600\GeV}{m_{\Q}}\right).
    \end{equation}

Subsequently, we decouple the production of $\Q$ from its decay into an invisible particle $\dm$ 
using the narrow width approximation (NWA), as implemented in \texttt{MadSpin} \cite{Artoisenet:2012st}.
For technical simplicity, we neglect spin correlation between the production and decay of $\Q$.
For $\stp$ and $\sigma$ decays, this has no impact due to their scalar nature.
For $\tp$ and $\go$ decays, this can impact the $p_T$ and $\eta$ distributions of $X$ and 
subleading jets associated with $\Q$'s decay, and hence the true $\met$ distribution.
However, we have checked that this approximation has a relatively small impact (up to 5\%) on our 
specific study since the $\met$ observable we use is built from visible objects, 
and is dominated by contributions from the hard radiation.

 More specifically, for each $\Q$, the particle nature of the DM particle $\dm$ is chosen in accordance with its underlying theory:
 For $\tp$, $\dm$ is a neutral scalar and $\tp$ are decayed to the two-body final state $\dm+\q$, where $\q$ is a light QCD quark.
 For $\stp$, $\dm$ is a neutral fermion and $\stp$ are decayed  via an off-shell top quark to the four-body final state, $\stp \to t^* \dm \to \dm+ b\qqbar'$.
 For $\go$, $\dm$ is a neutral fermion and $\go$ are decayed to the three-body final state $\dm+\qqbar'$. 
 We enforce a compressed mass spectrum by fixing a small mass gap between $\Q$ and $\dm$ to be $\Delta m = m_{\Q} - m_{\dm} = 20\; \mathrm{GeV}$.
 Since the mass gap is (relatively) small, the SM decay products of $\Q$ are forced to be (relatively) soft.
 Hence, the SM decay products of $\Q$ fail the criteria needed to be identified as the leading jet, 
further justifying our neglect of spin-correlation in the decays of $\tp$ and $\go$.
 Qualitatively and quantitatively, 
 our results are expected to be robust against varying $\Delta m$ factors of $2$ as this does not change substantially the likelihood of $\Q$ decay promptly, 
 nor the likelihood of its decay products being tagged as subleading jets in the event.
 For pathologically smaller $\Delta m_{\Q}$, the lifetime of $\Q$ can be extended beyond $ 100~\mu m$, but does not change the prompt monojet collider signature, nor our conclusions.
 For further discussion on displaced (monojet) signatures, see \cite{CMS-PAS-EXO-16-003,Aaboud:2017iio}.
 For hierarchically larger $\Delta m_{\Q}$, the SM decay products of $\Q$ become more energetic and the process transitions to a multi-jet+$\met$ collider signature.
 The signature associated with the latter scenario is outside the scope of our study and hence will not be discussed further.

 The choices of $\Q$, their relevant quantum numbers and decay path, and the corresponding 
 UFO library references are summarized in \tableref{tab:model}.

\subsection{Multi-Leg Matching and Merging Prescriptions}
\label{sec:SetupMatching}

The collider signature considered in this work is characterized by the presence of a high-$\pt$ jet 
recoiling against the $\QQbar$ system. 
It is thus necessary to include at least one QCD radiation at the matrix element level beyond what is modeled in 
Born-level $\QQbar$ production.
With presently available MC technology, this can be achieved in different ways and at various formal accuracies.
We now briefly describe the several prescription used in this study.

\begin{itemize}
\item {\bf \ul{Leading Order Multijet Merging}}: 
The LO multijet/multileg merging techniques \cite{Lonnblad:2001iq,Mangano:2006rw,Lonnblad:2012ng} 
outline how parton shower emissions can be augmented with full matrix elements.
The emissions are classified according to their hardness, i.e., $\pt$, and in terms of a dimensionful variable $\Qcut$. 
Emissions above a hardness threshold $\Qcut$ are described at LO accuracy using the appropriate matrix elements 
while  preserving the all-orders resummation accuracy of the parton shower below $\Qcut$.
In this work we use the  MLM scheme \cite{Mangano:2006rw} as implemented in \texttt{MadGraph5\_aMC@NLO} \texttt{2.6.0} (\mgamc)~\cite{Alwall:2014hca}. 
We take into account matrix element corrections to $\QQbar$ pair production  in association with up to two QCD partons.
While genuine $\mathcal{O}(\alpha_s)$ (and higher) corrections are included via this procedure,
the calculation remains formally LO accurate (LO+LL after parton showering) due to missing virtual corrections. Some of the earlier studies on 
monojet spectra with LO multijet merging technique using then available Madgraph/MadEvent v4 \cite{Alwall:2007st}, we 
refer~\cite{Alwall:2008va,Alwall:2008qv,LeCompte:2011fh,Dreiner:2012gx,Dreiner:2012sh}. 
\vskip 0.3cm
  
\item {\bf \ul{$\QQbar$ Production at Next-to-Leading Order with Parton Shower Matching}}: 
The accuracy of fixed order (FO) matrix element calculations at NLO can be combined with resummed parton showers at LL 
by means of NLO+PS matching techniques~\cite{Frixione:2002ik,Nason:2004rx,Frixione:2007vw}, such as the MC@NLO formalism \cite{Frixione:2002ik}.
In this approach, the LO matrix element for an extra hard, wide-angle QCD emission in the final state is naturally included as part of the $\mathcal{O}(\alpha_s)$
FO correction. Extra soft and/or collinear emissions enter through the parton shower.
Potential double counting of $\mathcal{O}(\alpha_s)$ soft/collinear contributions is avoided by the use of additional counter terms.
It is worth noting that leading-jet observables in this calculation are at most LO+LL accurate, but are nonetheless well-defined at all $\pt$.
\vskip 0.3cm
  
\item {\bf \ul{$\QQbarj$ Production at Next-to-Leading Order with Parton Shower Matching}}: 
In order to achieve NLO+LL accuracy for leading-jet observables, 
the above NLO+PS matching technique must be applied to the  $\QQbarj$ process. 
This requires explicitly introducing a regularizing $\pt$ selection on the leading jet at the matrix element level.
Since no Sudakov form factor is present for this jet, 
the $\pt$ selection must also be well above the Sudakov shoulder of the inclusive $\QQbar$-system
to ensure that the FO is perturbatively valid.
\vskip 0.3cm
  
\item {\bf \ul{Next-to-Leading Order Multijet Merging}}: 
The LO multijet merging technique described above can be extended to describe jet observables at NLO precision for jets above a hardness scale $Q_\mathrm{cut}$. 
Analogous to the LO case, matrix element merging with parton shower matching at next-to-leading order, known colloquially as MEPS@NLO,
is achievable by introducing additional all-orders/resummed Sudakov form factors for each NLO-accurate matrix element in consideration.
We use an extension of the Catani-Krauss-Kuhn-Webber (CKKW)~\cite{Catani:2001cc,Krauss:2002up,Schalicke:2005nv} merging formalism 
as implemented in \sherpa~by Refs.~\cite{Gehrmann:2012yg,Hoeche:2012yf}.
In the following, we employ MEPS@NLO multijet merging with up to one or two jets,
meaning that samples will contain up to two or three real radiations, respectively, beyond the lowest order process before parton showering.
As in the LO(+LL) case, while $\mathcal{O}(\alpha_s^2)$ corrections are present in this calculation, 
the final result remains formally NLO(+LL) accurate.

\end{itemize}
\subsection{Event Generation in MadGraph5\_aMC@NLO + Pythia8}
\label{sec:SetupMGPY}

Cross section calculations and event generation with accuracy up to NLO in QCD are handled
using \mgamc. For signal processes, we use ~the NLO-accurate UFO libraries described 
in Sec.~\ref{sec:SetupModels} and listed in Table.~\ref{tab:model} as inputs to \mgamc.
Within the \mgamc~framework, one-loop virtual corrections are evaluated numerically via \texttt{MadLoop}~\cite{Hirschi:2011pa} 
and matched with real emissions using the Frixione-Kunszt-Signer (FKS) subtraction formalism~\cite{Frixione:1995ms}, as implement by Ref.~\cite{Frederix:2009yq}.
Decays of $\Q$ are then handled at LO accuracy with \texttt{MadSpin}~\cite{Artoisenet:2012st}.
The central value $Q_0$ for the renormalization scale $Q_R$ and factorization scale $Q_F$ is set to
\begin{equation}
 Q_F,~Q_R = Q_0 \equiv \frac{H_T}{2}, 
 \;\text{with}\; H_T \equiv \sum_{k\in\{\g,\overset{(-)}{\q},\overset{(-)}{\Q}\}} \sqrt{m_k^2 + \vert \ptvec^k  \vert^2}
\end{equation}
where $H_T$ is the scalar sum of the transverse energy of the final state partons and $\Q$.

In \confirm{\sectionref{sec:Monojet}}, where we discuss experimental searches and sensitivity to $\Q$ in monojet searches,
we impose an analysis-level selection on the leading jet $(j_1)$ in an event. 
More specifically, we set the \texttt{ptj} variable in \mgamc's \texttt{run\_card.dat} file to \texttt{ptj}$=\ptjcut-100\;\gev$ (and jet radius $R=0.6$).
Then, to enhance yields at relatively high $\pt$, we generate events by binning the phase space in $\pt$. 
After preparing event samples for a particular $\ptjcut$, we apply higher $\ptjcut$ until the statistical uncertainty of the MC samples is no longer negligible.
At this point, we prepare another event sample based on $\ptjcut$ and apply the procedure iteratively. 
For the samples with the highest $\ptjcut$, we apply exponential biasing on $\ptj$ at the event-generation level to enhance the tail of jet $\pt$ distributions.

After the $\QQbar$ pair have been decayed, events are passed to \texttt{Pythia 8.2.26}~\cite{Sjostrand:2014zea} 
for parton showering and hadronization. We choose a shower starting scale $Q_S$ small enough 
such that light, colored final-state partons in the matrix element remain the hardest emissions in the full process, if the parton exists.
Namely, the parton shower is restricted to operate below the scale~\cite{Alwall:2014hca}
\begin{equation}
Q_S = Q_{S*} \approx \min \left[ \frac{H_T}{2}, \sqrt{d_*} \right].
\end{equation}
Here, $d_*$ is the minimum $d_{i,j}$ distance measure as calculated with the $k_T$ algorithm~\cite{Catani:1993hr,Ellis:1993tq} (for $R=1$)
over all  momentum recombinations of light, colored partons during the clustering phase in fixed order event generation. 
The events are essentially categorized by whether or not a hard $\mathcal{O}(\alpha_s)$ emission is present.

To quantify and estimate the size of missing, higher order QCD corrections, we compute the three-point scale-variation envelope. 
This is obtained in the usual fashion, i.e., by varying discretely and jointly
the factorization and renormalization scales $Q_R$ and $Q_F$ over the range,
\begin{equation}
 0.5\times Q_0 \leq Q_F,~Q_R \leq 2.0\times Q_0.
 \label{eq:qcdScaleUnc}
\end{equation}
Where necessary, we also consider the uncertainty associated with the parton shower starting scale.
We quantify this by discretely and independently computing the scale variation  over the range,
\begin{equation}
 \min \left[ 0.5 \times \frac{H_T}{2}, \sqrt{d_*} \right] \leq Q_S \leq \min \left[ 2.0 \times \frac{H_T}{2}, \sqrt{d_*} \right].
 \label{eq:showerScaleUnc}
\end{equation}

\subsection{Event Generation in Sherpa}\label{sec:SetupSherpa}

For cross-validation of the fermionic top quark partner $\tp$, we employ~\sherpa~\texttt{2.2.4}~\cite{Gleisberg:2008ta}. 
At LO in QCD, arbitrary BSM models can be simulated through \sherpa's generic UFO model~\cite{Degrande:2011ua} interface \cite{Hoche:2014kca}.
At NLO, processes involving $\tp$ can be simulated by slightly modifying the default SM model file, 
and setting the top quark mass to the mass of $\tp$.
For the decay into a scalar dark matter particle, an additional decay vertex is added using the methodology of Ref.~\cite{Hoche:2014kca}.
Tree-level matrix elements of the calculation are provided by \sherpa's in-house matrix element generators \texttt{AMEGIC} \cite{Krauss:2001iv} 
and \texttt{COMIX} \cite{Gleisberg:2008fv}. 
One-loop amplitudes are treated by interfacing with \texttt{OpenLoops} \cite{Cascioli:2011va}. 
Parton shower-matching is performed according to the MC@NLO formalism~\cite{Frixione:2002ik,Gleisberg:2007md,Hoche:2012wh},
using \sherpa's Catani-Seymour subtraction-based shower procedure~\cite{Catani:1996vz,Schumann:2007mg}.

To study potential improvements to modeling fermionic top quark partners, we also employ multijet merging at NLO in QCD with \sherpa.
To account for additional high-$\pt$ QCD emissions at the matrix-element level, beyond what is already present in inclusive $\tptp$ production at NLO, 
we include the NLO-accurate matrix elements for the processes
\begin{equation}
 \pptptpj \quad\text{and}\quad \pptptpjj.
\end{equation}
We merge these samples with the fully inclusive $\pptptp$ sample following the MEPS@NLO prescription \cite{Gehrmann:2012yg,Hoeche:2012yf}. 
In this scheme, the nominal values for the factorization, renormalization, and parton shower scale are determined through a backward clustering procedure 
that maps higher multiplicity configurations to a $2\to 2$ configuration.
We set all scales to the invariant mass of the top partner pair. 
As a nominal value for the merging scale $\Qcut$, we set $\Qcut = 120\;\mathrm{GeV}$.
For parton showering we employ one of \sherpa's dipole showers, which
is published in \cite{Schumann:2007mg}.

In addition to MEPS@NLO merged samples we also simulate $\pptptpj$ at NLO+PS with \sherpa.
We do not add matrix element corrections to these samples beyond what is already present at NLO+PS.
For the generation of the $\pptptpj$ process, we use the scale schemes,
\begin{equation}
  Q_F = Q_R =  m(\tptp)\, \quad\text{and}\quad Q_S = \pt(\tptp)\,,
\end{equation}
where $m(\tptp)$ and $p_T(\tptp)$ denote the $(\tptp)$-system's invariant mass and transverse momentum respectively.

\subsection{Detector simulation and object reconstruction}\label{sec:SetupReco}

For fast detector simulation, we use \texttt{Delphes 3.3.3} \cite{deFavereau:2013fsa} with the default ATLAS card. 
Jets are constructed from calorimeter tower elements using \texttt{Fastjet  3.2.1}~\cite{Cacciari:2011ma},
according to the anti-$k_T$ jet clustering algorithm ~\cite{Cacciari:2008gp} with jet radius $R=0.4$.
Analysis-quality jets are required to satisfy the fiducial and kinematic criteria,
\begin{equation}
 \pt^j > 20\GeV \quad\text{and}\quad \vert \eta^j\vert < 4.5.
\end{equation}
Events are then accepted or rejected based the monojet selection criteria discussed in Sec.~\ref{sec:Monojet}.

\subsection{Standard Model Inputs}\label{sec:SetupSM}

We assume $n_f=5$ active/massless quark flavors and a diagonal Cabbibo-Kobayashi-Maskawa (CKM)
quark mixing matrix with unit entries. Relevant SM inputs used in our study include,
\begin{eqnarray}
 m_t(\rm pole) = 173.3\GeV 	\quad\text{and}\quad	\alpha^{\rm \overline{MS}}(M_{Z})= 0.118.
 \label{eq:smInputs}
\end{eqnarray}
The evolution of parton distribution functions (PDFs) and the strong coupling constant $\alpha_s(\mu_R)$ are extracted using the 
\texttt{LHAPDF 6.1.6}~\cite{Buckley:2014ana} libraries. 
As discussed in Sec.~\ref{sec:SetupMatching}, 
LO (NLO) multijet merging with two (one) additional partons accounts for new kinematic channels and configurations that first arise at NNLO.
(In principle, all one would need to achieve NNLO accuracy are the missing two-loop virtual corrections.) 
However, such $\mathcal{O}(\alpha_s^2)$ contributions are already accounted for in the normalizations of NLO PDFs.
Hence, to minimize potential double counting of initial-state contributions, we use the NNPDF 3.0 NNLO PDF set~(\texttt{lhaid=261000})~\cite{Ball:2014uwa}
for all signal process calculations. For the LO SM $\ppZj$ calculation in \sectionref{sec:Monojet}, we use the NNPDF 3.0 LO PDF.


\section{Monojet Searches at the HL- and HE-LHC}\label{sec:Monojet}
\label{monojet}

At hadron colliders, the term ``monojets'' represents a broad class of sensitive collider signatures and search strategies
that assume varying degrees of particle multiplicity and inclusiveness.
In this section, we consider specifically the inclusive monojet signature, as implemented by ATLAS during Run II of the LHC's operations
after collecting $\mathcal{L}=$36.1~fb$^{-1}$ of integrated luminosity at $\sqrt{s}=13$ TeV~\cite{atlas-conf-2017-060}.
After discussing various sources of experimental and theoretical uncertainties,
we report the observed and expected sensitivity of the channel at current and proposed $pp$ colliders 
using the $\cls$ modified frequentist approach~\cite{Read:2002hq}.
The model-independent limits derived in Sec.~\ref{sec:monojetLimits} are then applied in Sec.~\ref{sec:heavyQLimits} 
to the heavy colored particles $\Q$ described in Sec.~\ref{sec:SetupModels}.
The ability to determine and distinguish principle properties of $\Q$ is then discussed in Sec.~\ref{sec:ident}.

\subsection{Monojet Searches, LHC Data, and Model-Independent Limits}\label{sec:monojetLimits}

\begin{table*}[!t]
\caption{The expected number of SM background events and associated errors for the inclusive mode signal regions IM1-IM10 
as defined by the ATLAS experiment in Ref.~\cite{atlas-conf-2017-060}.
}
\label{tab:stat}
\begin{center}
\begin{tabular}{rcrrr}
\hline\noalign{\smallskip}
\multirow{2}{*}{IM} & \multirow{2}{*}{$\etmiss$ [GeV]} & Expected SM Events with &  \multirow{2}{*}{Statistical Error} & \multirow{2}{*}{Total Error} \cr
		    & 				       & Total (Stat.+Sys.+Th.) Error   &					 & \cr
\noalign{\smallskip}\hline\noalign{\smallskip}
1  &$>250\mo$& $245900	\pm 5800$     	&   496 (0.20\%)  		& 2.3 \% \cr
2  &$>300\mo$&$138000  	\pm 3400$     	&   371 (0.27\%)  		& 2.5 \% \cr
3  &$>350\mo$& $73000  	\pm 1900$     	&   270 (0.37\%)  		& 2.6 \% \cr
4  &$>400\mo$&$39900   	\pm 1000 $    	&   200 (0.50\%)  		& 2.5 \% \cr
5  &$>500\mo$&$12720   	\pm 340\mo$	&  113  (0.89\%)  		& 2.6 \%\cr
6  &$>600\mo$&$4680    	\pm  160\mo$    &  \hphantom{0}68 (1.46\%)  	& 3.4 \%\cr
7  &$>700\mo$&$2017    	\pm 90\moo$	&  \hphantom{0}45 (2.23\%)  	& 4.4 \% \cr
8  &$>800\mo$&$908     	\pm 55\moo$    	&  \hphantom{0}30 (3.32\%)  	& 6.1 \% \cr
9  &$>900\mo$&$464     	\pm 34\moo$     &  \hphantom{0}22 (4.64\%)  	& 7.3 \% \cr
10 &$>1000$&$238    	\pm 23\moo$     &  \hphantom{0}15 (6.48\%)  	& 9.7 \% \cr
\noalign{\smallskip}\hline
\end{tabular}
\end{center}
\end{table*}

The ATLAS and CMS collaborations have both reported on their search of early Run II data 
for anomalous events with significant transverse momentum imbalance and at least one energetic jet~\cite{atlas-conf-2017-060,sirunyan:2017jix}.
For the case under consideration~\cite{atlas-conf-2017-060}, the ATLAS collaboration has investigated two overlapping signal regions, 
categorized as exclusive modes (EM) and inclusive modes (IM), based on various $\met$ thresholds spanning $\met=250$ GeV to 1 TeV. 
The EM signal regions are defined in terms of $\met$ binning. 
For example: signal region EM1 (EM5) selects for events with $\met$ satisfying
250 GeV~$< \met < $~300 GeV~(500 GeV~$< \met < $~600 GeV). The IM signal regions are defined in terms of minimum $\met$ selections.
For example: signal region IM1 (IM5) selects for events with $\met$ satisfying $\met >$ 250 (500) GeV.
Additionally, events are required to satisfy the following selection criteria:  
\begin{itemize}
\item At least one analysis-level jet with $p_T>250$~GeV and $\vert\eta\vert<2.4$.
\item A maximum of four analysis-level jets satisfying $p_T>30$~GeV and $\vert \eta\vert<2.8$.
\item An azimuthal separation of $\Delta\phi(jet, \met) >0.4$ for each analysis-level jet and the $\met$ vector.
\end{itemize}
For the remainder of this study, we focus on the IM monojet signal regions.

In \tableref{tab:stat}, we display the expected number of SM (background) events passing all selection criteria 
with uncertainties (statistical and total) in each of the inclusive mode signal regions (IM1-IM10), as reported by ATLAS~\cite{atlas-conf-2017-060}.
Non-statistical errors include both experimental and theoretical uncertainties.
Sources of systematic uncertainty  estimated in Ref.~\cite{atlas-conf-2017-060} include:
dependencies on parton shower and PDF modeling, which span $\pm$0.7\%  to $\pm$0.8\%;
uncertainties in jet energy and $\met$ scales, which range $\pm$0.5\% (IM1) to $\pm$5.3\% (IM10); 
jet quality and pileup descriptions additionally provide uncertainties ranging $\pm$0.8\% to $\pm$1.8\%; 
for more detailed discussions,  see Ref.~\cite{atlas-conf-2017-060}. 
We note that the total errors for IM1-5 are nearly flat, with 2.2\% to  2.6\%, indicating that these signal regions' uncertainties are systematics dominated.
Statistical and systematic uncertainties are much larger in the higher $\met$ regions.
However, with the HL phase of LHC, one expects to collect significantly more data
that will correspondingly reduce statistical errors for the high-$\met$ regions.
Additionally, the analysis' control sample will also increase during the HL run, therefore also reducing systematic uncertainties.
Thus, one anticipates that total uncertainties will shrink for future inclusive monojet searches at LHC.

\begin{table*}[!t]
\caption{Model-independent 95\% CL upper limit on the visible cross section for each inclusive mode (IM) signal region,
after $\mathcal{L}=$36.1~fb$^{-1}$ of data at $\sqrt{s} = 13$ TeV, as reported by Ref.~\cite{atlas-conf-2017-060},
and the estimated $\cls$ limit assuming $\mathcal{L}=$3~ab$^{-1}$.
}
\label{tab:limit}
\begin{center}
\begin{tabular}{crrrr}
\hline\noalign{\smallskip}
\multirow{2}{*}{IM} &  \multirow{2}{*}{Observed limit [fb]} & \multirow{2}{*}{Expected limit [fb]}  
		    & \multicolumn{2}{c}{Scaled limit [$\mathrm{fb}$] for $\mathcal{L}=3$ ab$^{-1}$}  \cr
\cmidrule(rl){4-5}   &        &   &  2.5\% Syst. Error    &  1\% Syst. Error   \cr
\noalign{\smallskip}\hline\noalign{\smallskip}
1  &  531\hphantom{.}\mo     	&     324\hphantom{.}\mo     &  333\hphantom{.}\mo  & 133\hphantom{.}\mo\cr
2  &  330\hphantom{.}\mo     	&     194\hphantom{.}\mo     &  187\hphantom{.}\mo  &  75\hphantom{.}\mo \cr
3  &  188\hphantom{.}\mo     	&     111\hphantom{.}\mo     &  99\hphantom{.}\mo   &  39\hphantom{.}\mo \cr
4  &  93\hphantom{.}\mo      	&      58\hphantom{.}\mo     &  54\hphantom{.}\mo   & 22\hphantom{.}\mo \cr
5  &  43\hphantom{.}\mo      	&      21\hphantom{.}\mo     &  17\hphantom{.}\mo   &   6.9\cr
6  &  19\hphantom{.}\mo		&  	9.8  &  6.4  &   2.6\cr
7  &  7.7     			&     5.7    &  2.8  &  1.1\cr
8  &  4.9     			&     3.4    &  1.2  &  0.5\cr
9  &  2.2     			&     2.1    &  0.6  &  0.3\cr
10 &  1.6     			&     1.5    &  0.3  &  0.2\cr
\noalign{\smallskip}\hline
\end{tabular}
\end{center}
\end{table*}

The non-observation of data deviating significantly from SM predictions enables one to set 
model-independent upper limits on the production cross section of new particles. 
In the second and third columns of \tableref{tab:limit}, we tabulate the expected and observed limits at the 95\% 
confidence level (CL) on the visible cross section\footnote{Explicitly, the visible cross section is 
defined as the product of total cross section, acceptance, and efficiency.}, respectively, for each IM signal 
region, as reported in~\cite{atlas-conf-2017-060} with $\mathcal{L}=$36.1~fb$^{-1}$ of data.
To address the prospect of the HL-LHC with $\mathcal{L} = 3$ ab$^{-1}$ at $\sqrt{s}$ = 13 TeV, 
we calculate the expected upper limits by scaling the number of events at the 13 TeV run of LHC.
We choose two values of the total systematic uncertainty, namely $\delta_{\rm Sys.}=$2.5\% and 1\%.
The former is a pessimistic assertion that systematic uncertainties will not be reduced beyond present levels (see~\tableref{tab:stat}),
even after 15-20 years of LHC operations. The latter is an optimistic, but benchmark, assumption.
The likelihoods of background only and signal-plus-background hypotheses are set as Gaussian, with a standard deviation set to the total uncertainty.
We have checked that our likelihoods are in agreement with those reported by Ref.~\cite{atlas-conf-2017-060} for $\mathcal{L} = 36.1$ fb$^{-1}$.
We report the scaled limits in the last two columns of \tableref{tab:limit}. 
We now discuss the impact of these limits on the production of $\QQbar$ pairs in $pp$ collisions.

\subsection{HL- and HE-LHC Sensitivity to Heavy Colored Particles $\Q$}\label{sec:heavyQLimits}


\begin{figure*}[t]
\begin{center}
\includegraphics[width=0.40\textwidth]{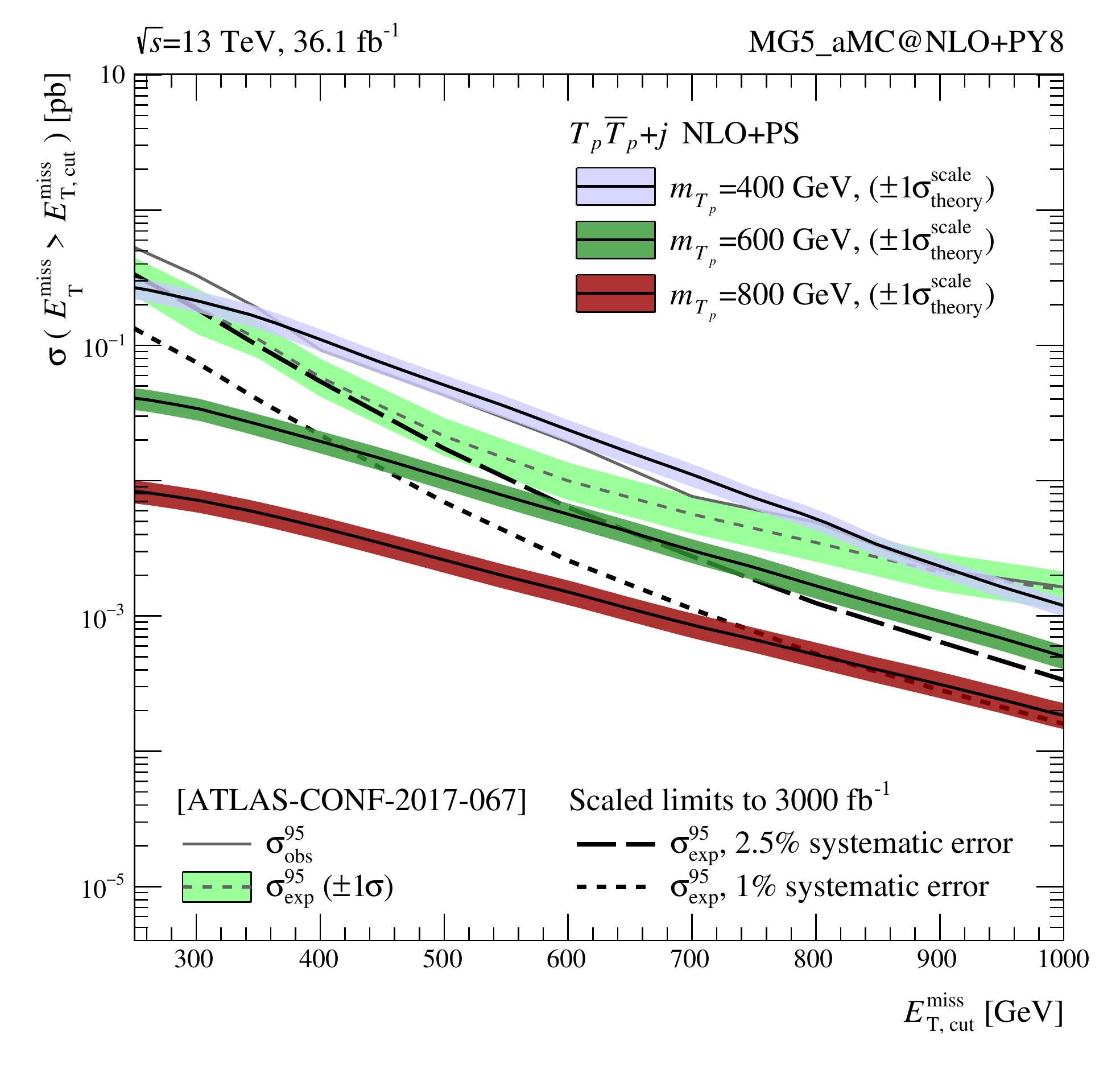}	
\includegraphics[width=0.40\textwidth]{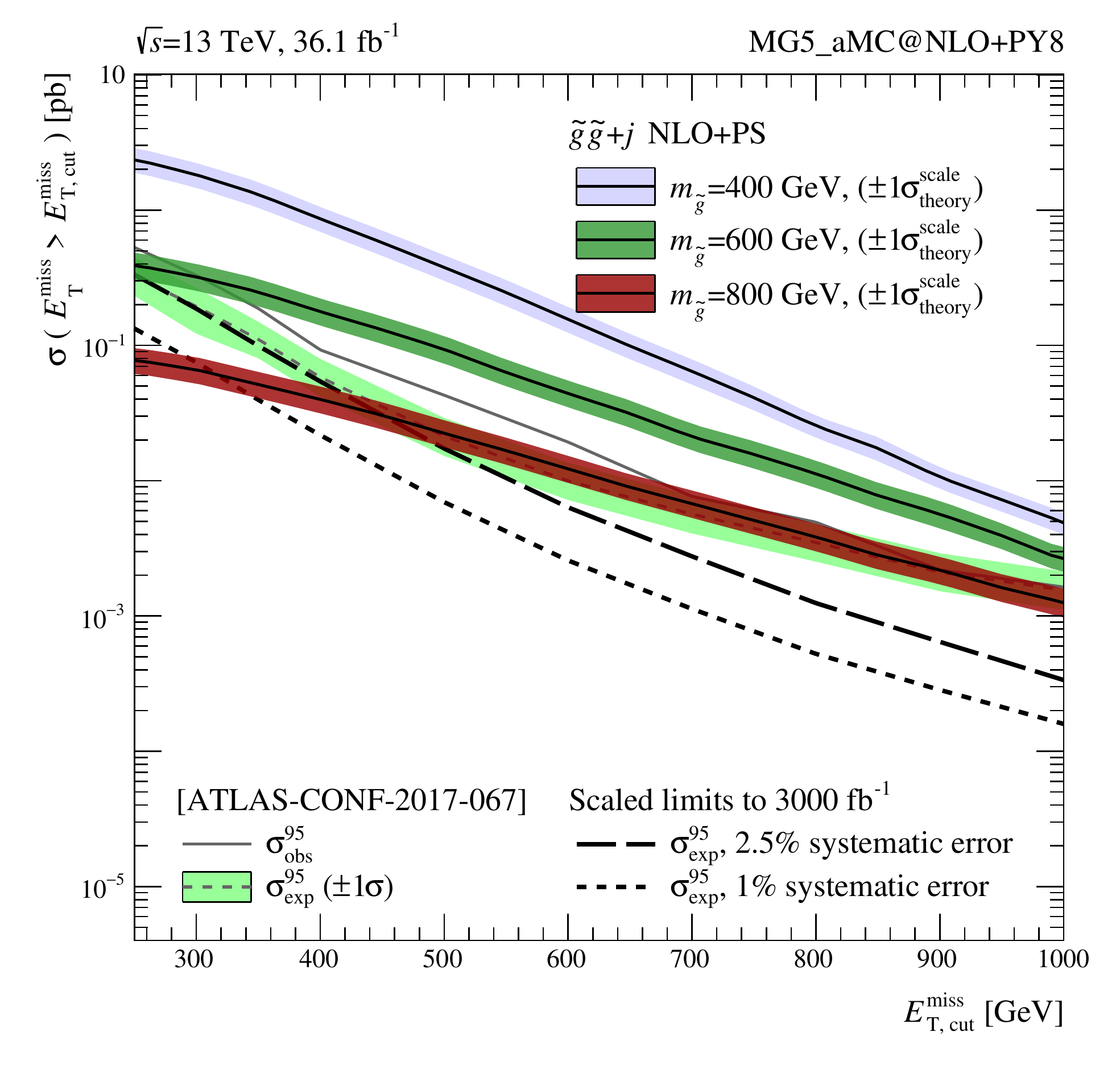}	
\includegraphics[width=0.40\textwidth]{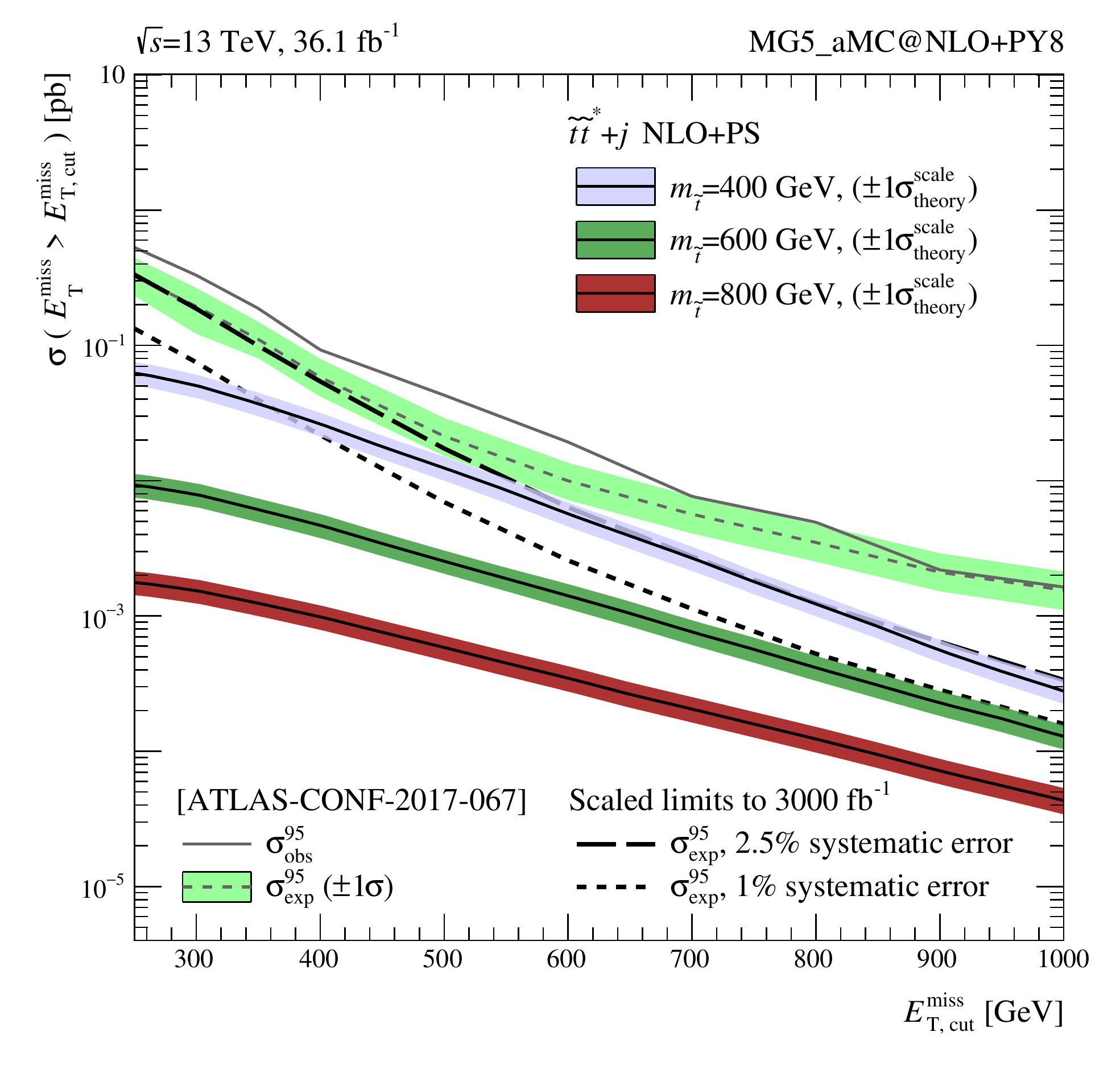}
\end{center}
\caption{
The $\ppQQbarj$ cross section as a function of minimum $\met$ after the experimental selection criteria at 13 TeV,
for $\Q=\tp,~\go$, and $\stp$, with current 95\% CL$_s$ limits after $\mathcal{L}=36.1$ fb$^{-1}$ of data at the 13 TeV LHC. 
Also shown is the estimated sensitivity with $\mathcal{L}=3$ ab$^{-1}$, assuming $\delta_{\rm Syst.}=$2.5\% and 1\% systematical errors.
}\label{fig:comp13TeV}
\end{figure*}

We now compare and apply the model-independent upper limits on the cross sections derived in the previous section to the production of $\QQbar$ pairs at the LHC.
In \figref{fig:comp13TeV}, we show the model-independent 95\% CL upper limits along with NLO+PS-accurate cross section for $\Q$ produced 
in association with a hard jet at the matrix-element level, viz. $\QQbar+j$ at NLO. 
(For details of our computational setup, see \sectionref{sec:numeric}.) We overlay FO scale uncertainty, computed according to 
\eqref{eq:qcdScaleUnc}. For all cases of $\Q$, we find that the scale uncertainty at NLO is a 
dominant source of uncertainty and hence take it as a representative measure of the total uncertainty.
A dedicated and in-depth discussion of this uncertainty is given in \sectionref{sec:theoUncertainty}.

From inclusive monojet searches at $\sqrt{s} = 13$ TeV with $\mathcal{L}=36.1 \ifb$ of data, 
we find that the lower limits on $\Q$ masses stand at around 
$m_{\tp}=$ 400 GeV for the fermionic top partner and $m_{\go}=$ 600 GeV for the gluino,
while no constraint on stop masses is found within the range under consideration.
We observe that in the high-$p_T$ bins both the systematic and statistical experimental uncertainties play a crucial role.
As argued in the previous section, one expects sensitivity to improve at the HL-LHC due to a much larger dataset,
leading to better control on both uncertainties.
From the scaled limits, we find that fermionic top partners with masses $m_{\tp}\lesssim$ 800 GeV, gluinos with $m_{\go}\lesssim$ 1000 GeV, 
and stops with masses $m_{\stp}\lesssim$ 600 GeV, in a compressed spectrum scenario, can be excluded at 13 TeV 
with $\mathcal{L}=3$ ab$^{-1}$, using the inclusive monojet signature.


Along with more data, a possibility that can greatly push the sensitivity to heavy colored particles is increasing the beam energy of the LHC itself.
Presently, community discussions are underway on upgrading the LHC's magnet system to handle a center-of-mass energy up to $\sqrt{s}=27$~TeV~\cite{conf27tev}. 
In light of this prospect, we briefly investigate the impact of a higher beam energy on the production of $\Q$ and the SM $Z+\mathrm{jets}$ background, i.e.,
the dominant background of the monojet signature~\cite{atlas-conf-2017-060}, and estimate the experimental reach of such a collider.

\begin{table}
\caption{The LO $Z+j$ cross section (pb) for representative $\ptgen$ (GeV) at $\sqrt{s}=13$, 14, and 27~TeV.  }
\label{tab:zjlo}
\begin{center}
   \begin{tabular}{rlll}
\hline\noalign{\smallskip}
\multirow{2}{*}{$\ptgen$ (GeV)}& \multicolumn{3}{c}{$pp\rightarrow Zj$ cross section (pb)} \\
     & \multicolumn{1}{c}{13~TeV} & \multicolumn{1}{c}{14~TeV} & \multicolumn{1}{c}{27~TeV} \cr
\noalign{\smallskip}\hline\noalign{\smallskip}
300   &     9.40     			& 11.1    			& 41.9\cr
400   &     2.59     			& \hphantom{0}3.11    		& 13.4\cr
500   &     0.889    			& \hphantom{0}1.09    		& \hphantom{0}5.22 \cr
600   &     0.350    			& \hphantom{0}0.438   		& \hphantom{0}2.36\cr
800   &     7.18 $\times 10^{-2}$  	& 9.33 $\times 10^{-2}$   	& \hphantom{0}0.637\cr
1000  &     1.85  $\times 10^{-2}$ 	& 2.49  $\times 10^{-2}$  	& \hphantom{0}0.214  \cr
1200  &     5.46  $\times 10^{-3}$ & 7.63  $\times 10^{-3}$  & 8.44 $\times 10^{-2}$ \cr
1400  &     1.77  $\times 10^{-3}$ & 2.59  $\times 10^{-3}$  & 3.69 $\times 10^{-2}$ \cr
1600  &     6.03 $\times 10^{-4}$ & 9.29 $\times 10^{-4}$  & 1.71 $\times 10^{-2}$ \cr
\noalign{\smallskip}\hline
 \end{tabular}
\end{center}
\end{table} 

Before this, however, we shortly digress to describe our modeling of the SM $Z+$jets background. In particular, we 
note that while the signal processes are consistently determined at NLO in QCD with parton shower matching, we 
consider the $Z+$jets background only at LO with an experimentally determined normalization factor. 
For our purposes, we believe this provides a sufficiently reliable description of the SM backgrounds after selection cuts.

The motivation comes precisely from the fact that proper modeling of the SM EW boson+jets background for 
monojet searches is highly nontrivial~\cite{Lindert:2017olm}, particularly in comparison to simulating the 
inclusive $W/Z+$jets process. The technical difficulty is due, in part, to strong phase space restrictions (cuts) 
on the final-state jets (see above Sec.~\ref{sec:monojetLimits} for the list of cuts). Present implementations of 
the MC@NLO formalism into general purpose event generators require that one integrates over the entire phase space 
of additional real radiation at NLO to ensure infrared pole cancellation. In the present case, this renders event 
generation at NLO inefficient. For example: the $pp\to Z+j$ process with $p_T^j \gtrsim 250-1000$ GeV is 
known~\cite{Rubin:2010xp,Frederix:2018nkq} to have giant QCD corrections stemming from the opening of new 
kinematic configurations. In this instance, large corrections are driven by the high-$p_T$ dijet process with 
a relatively soft $Z$ emission off a final-state quark, a configuration that would otherwise fail 
the $\Delta\phi(jet,\met)$ and minimum $\met$ selection criteria.  Moreover, after careful consideration, we 
find that the EW boson + jets background, that survives the selection analysis, is dominated by Born-like 
configurations. For example: after cuts, partonic channels such as $qg \to qZ$ contribute much more to the $pp\to Z+$jets background than channels like $qq\to qqZ$.

Once selection criteria have been applied, 
the difference then between our background modeling and a much more precise determination, e.g., Ref.~\cite{Lindert:2017olm}, is largely an overall normalization. 
This does not necessarily hold true when taking into account EW corrections at NLO and beyond.
However, such corrections are beyond the claimed accuracy of our work and we refer readers to Refs.~\cite{Rubin:2010xp,Frederix:2018nkq} for thorough discussions. 
For representative minimum $\met$ choices, 
we have checked at 14 TeV that the ratio $(K^{\rm NLO})$ of the $pp\to Z+j$ rates at NLO+PS and LO+PS after selection cuts are applied is roughly a constant $K^{\rm NLO} \sim 1.2$.
This is consistent with the size of finite virtual corrections at NLO in QCD to the DY process~\cite{Altarelli:1979ub}, 
and supports our arguments that the residual EW+jets background exhibits Born-like kinematics.
Now, to achieve a reliable normalization of the EW+jets background, we scale the generator-level $pp\to Z+j$ cross section at LO by (approximately) a factor 1/20 
so that the ratio of the curve to the signal cross section is normalized with respect to the post-event selection limit in \figref{fig:comp13TeV}.
For $\sqrt{s}=14$ and 27 TeV, this is additionally scaled by the cross section ratios 
$\sigma(\sqrt{s}=14\TeV)/\sigma(\sqrt{s}=13\TeV)$ and $\sigma(\sqrt{s}=27\TeV)/\sigma(\sqrt{s}=13\TeV)$, respectively, to account for the increase in parton luminosity.
In Table \ref{tab:zjlo}, we list the cross sections for the $Z+j$ process at $\sqrt{s}=13,~14,$ and 27 TeV for representative $\ptgen$ selections.
We stress that this procedure is only an estimation of the SM background; 
we do not advocate that this is a suitable replacement for full NLO+PS (or more accurate) computations in experimental searches.

For representative masses, $m_{\Q}$, we show in~\figref{fig:xsec14} the cross sections for the $pp\to \QQbar+j$ and 
(normalized) $Z+j$ processes as a function of the leading jet $p_T$.
More specifically, the cross sections are calculated as a function of a generator-level $p_T$ threshold $(\ptgen)$ on the light jet. 
In the lower panel, we show the 27 TeV-to-14 TeV cross section ratios.
There, one sees that the production cross section of $\Q$ increases faster than the SM background with increasing center-of-mass energy.
The enhancement follows from the well-documented~\cite{Martin:2009iq,Arkani-Hamed:2015vfh,Mangano:2016jyj} growth in PDF luminosities 
for fixed partonic mass scales but increasing collider beam energy.
Quantitatively, for $\ptgen=1$~TeV, the $Z+j$ cross section increases by a factor of 10 with respect to the change of $\sqrt{s}$,
while 
the fermionic top partner cross section for $m_{\tp} = 600~(800)$~GeV increases by approximately $24\times~(28\times)$,
the gluino rate for $m_{\go}=900$ GeV by $40\times$, 
and the stop rate for $m_{\stp}=400$ GeV by $20\times$.
Although other sources of SM backgrounds for the monojet signature have been presently neglected, 
the signal over background ratio $(S/N)$ still increases significantly at higher collider energies due to the larger luminosity enhancement for the 
signal process than dominant SM backgrounds.
Subsequently, the HE-LHC enables ones to investigate parameter regions that are not accessible at the LHC.
We do emphasize, however, that $S/N$ ratios can change drastically if additional information is provided to enhance 
the separation of the signal events from the backgrounds.
For example: proposals exist on how to utilize soft leptons, jets, and displaced vertices associated with decays of $\Q$
that can further reduce SM background rates~\cite{Chakraborty:2016qim,Nagata:2017gci,Chakraborty:2017kjq}.

As shown in \figref{fig:comp27TeV}, once the scaled limits for the SM backgrounds are determined, 
we can compare the predicted cross sections for $p p \rightarrow \Q \Qbar + j$ NLO process and estimate the expected reach at the 27 TeV HE-LHC.
We find that with $\mathcal{L}=3-15$ ab$^{-1}$, one is sensitive to compressed spectra scenarios featuring
fermionic top partners with masses $m_{\tp}\lesssim1100$ GeV,
gluinos with masses $m_{\go}\lesssim1800$ GeV,
and stops with masses $m_{\stp}\lesssim600$ GeV.

\begin{figure}
\begin{center}
\includegraphics[trim={0.5cm 0 0 0},clip,width=0.7\textwidth]{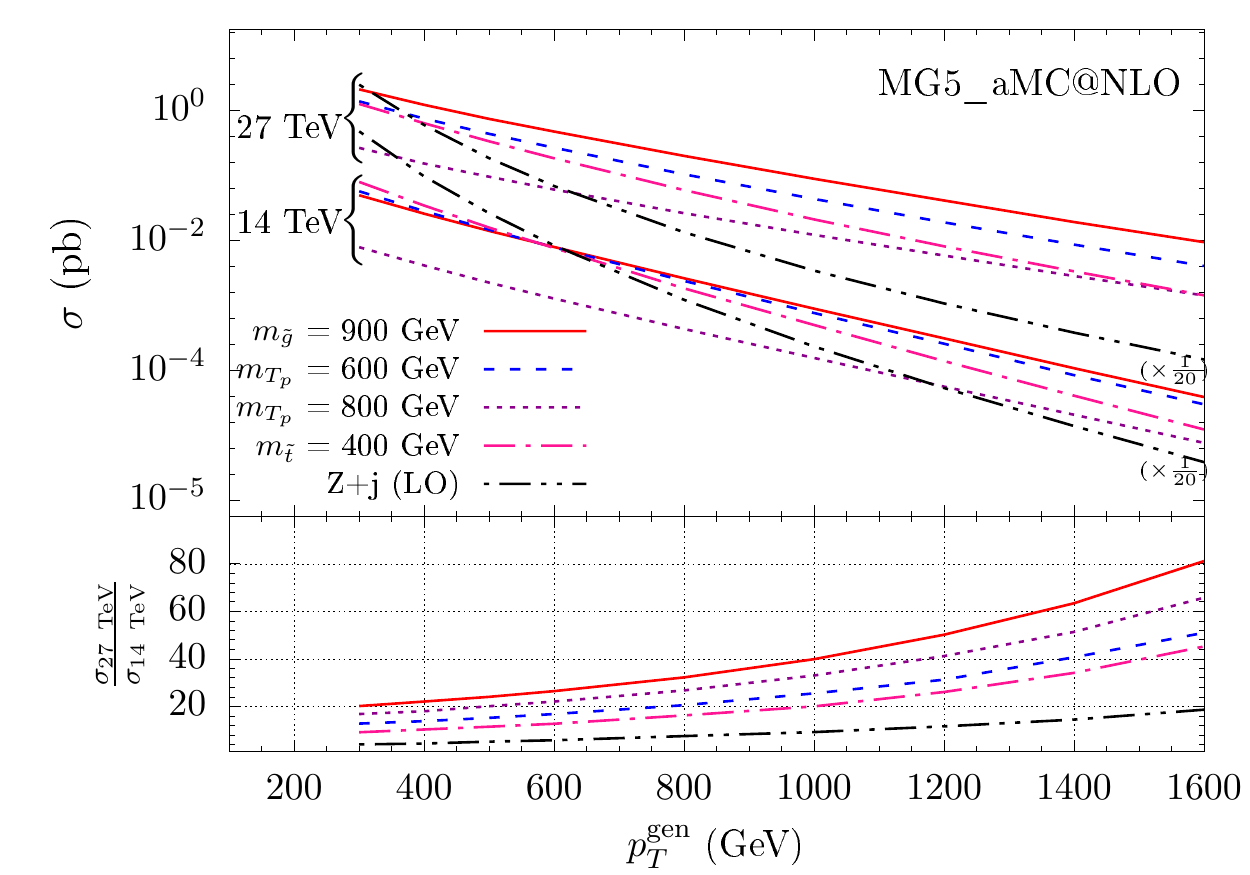}
\caption{Upper: Cross sections of  $\ppQQbarj$ process as a function of $\ptgen$ for various $\Q$ at $\sqrt{s}=14$~TeV and 27~TeV.
Lower: The ratio of the cross sections at 27 and 14 TeV.
}
\label{fig:xsec14}
\end{center}
\end{figure}

We end this discussion by providing an estimate of the anticipated sensitivity of new colored particles $\Q$ at the 27 TeV HE-LHC. 
Here, we assume that the SM background is dominated by the $Z+j$ process
and simply scale the model-independent 95\% CL upper limit at $\sqrt{s}=13$ TeV by the 27-to-13 TeV production cross section ratio.
In other words, the SM background cross section at a given $\sqrt{s}$ and $\metcut$, denoted as $\sigma ({\sqrt{s}}; {\metcut}) $, 
is obtained from the relation \hfill
\begin{equation}
\sigma ({\sqrt{s}}; {\metcut}') = \sigma ({13\;\mathrm{TeV}}; \metcut) 
\cdot \frac{\sigma_{pp \rightarrow Zj}(\sqrt{s};\ptgen = {\metcut}' )}{\sigma_{pp\rightarrow Zj}({13\;\mathrm{TeV}};\ptgen = \metcut )}.
\end{equation}
In the above, $\sigma_{pp \rightarrow Zj}(\sqrt{s};\ptgen)$ is the LO $Z+j$ cross section with $\ptgen$ at a collider energy $\sqrt{s}$.
We further assume that the detector acceptance and efficiencies are the same at 13 and 27 TeV.
This assumption is not as strong as one may anticipate in more general circumstances. 
The HE-LHC project proposes to refit, replace, and/or upgrade the current LHC magnet system and detector experiments.
As the detector experiment caverns themselves cannot physically grow, one is forced to adopt a detector fiducial volume at 27 TeV 
that is largely unchanged from 13 TeV.
Similarly, we also assume systematic uncertainties of 2.5\% and 1\%, the same considered in Sec.~\ref{sec:monojetLimits}.


\begin{figure*}
  \centering
  \begin{subfigure}{.49\textwidth}
  \includegraphics[width=\textwidth]{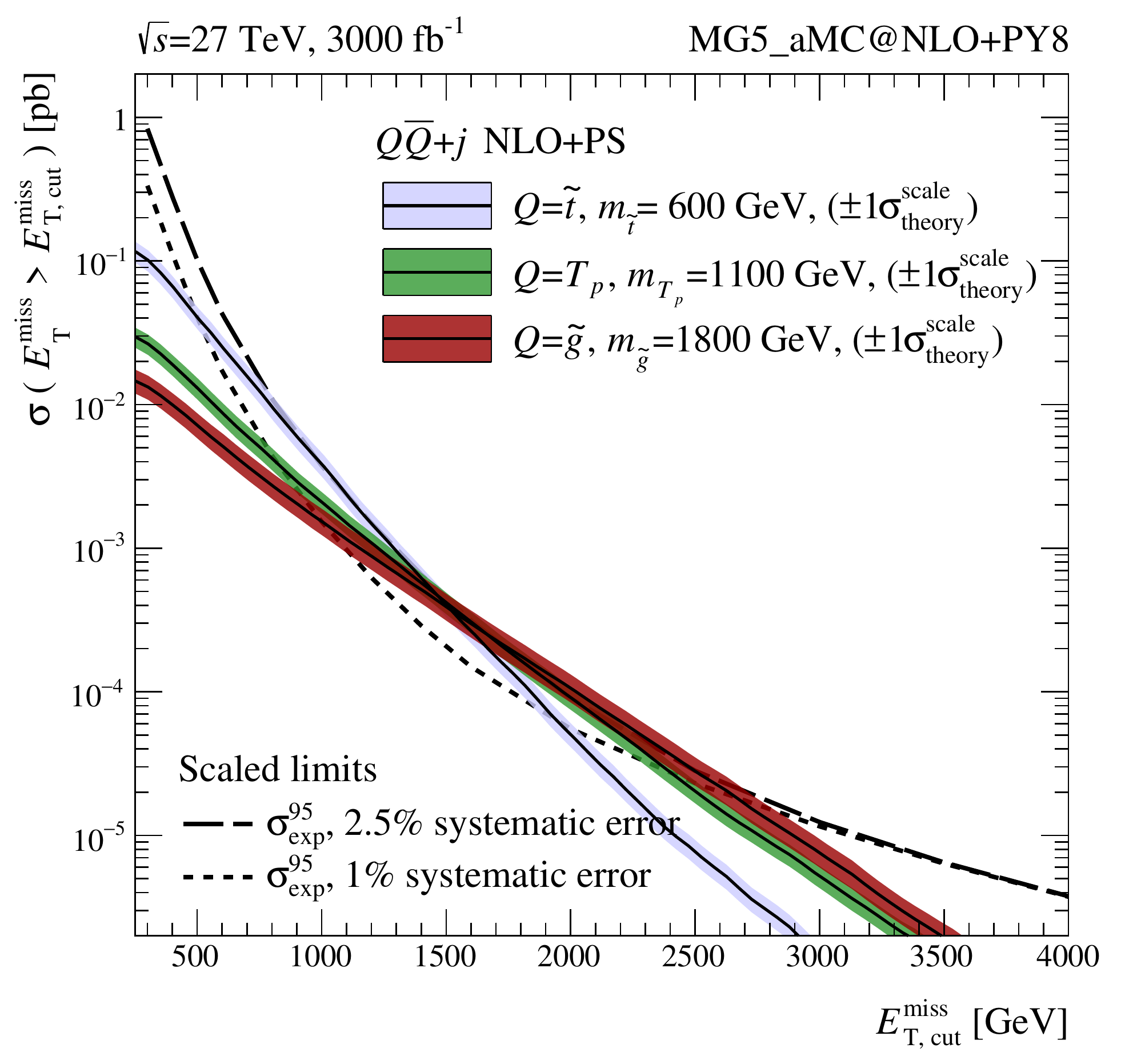} \caption{} 
  \end{subfigure}
  \hfill
  \begin{subfigure}{.49\textwidth}
  \includegraphics[width=\textwidth]{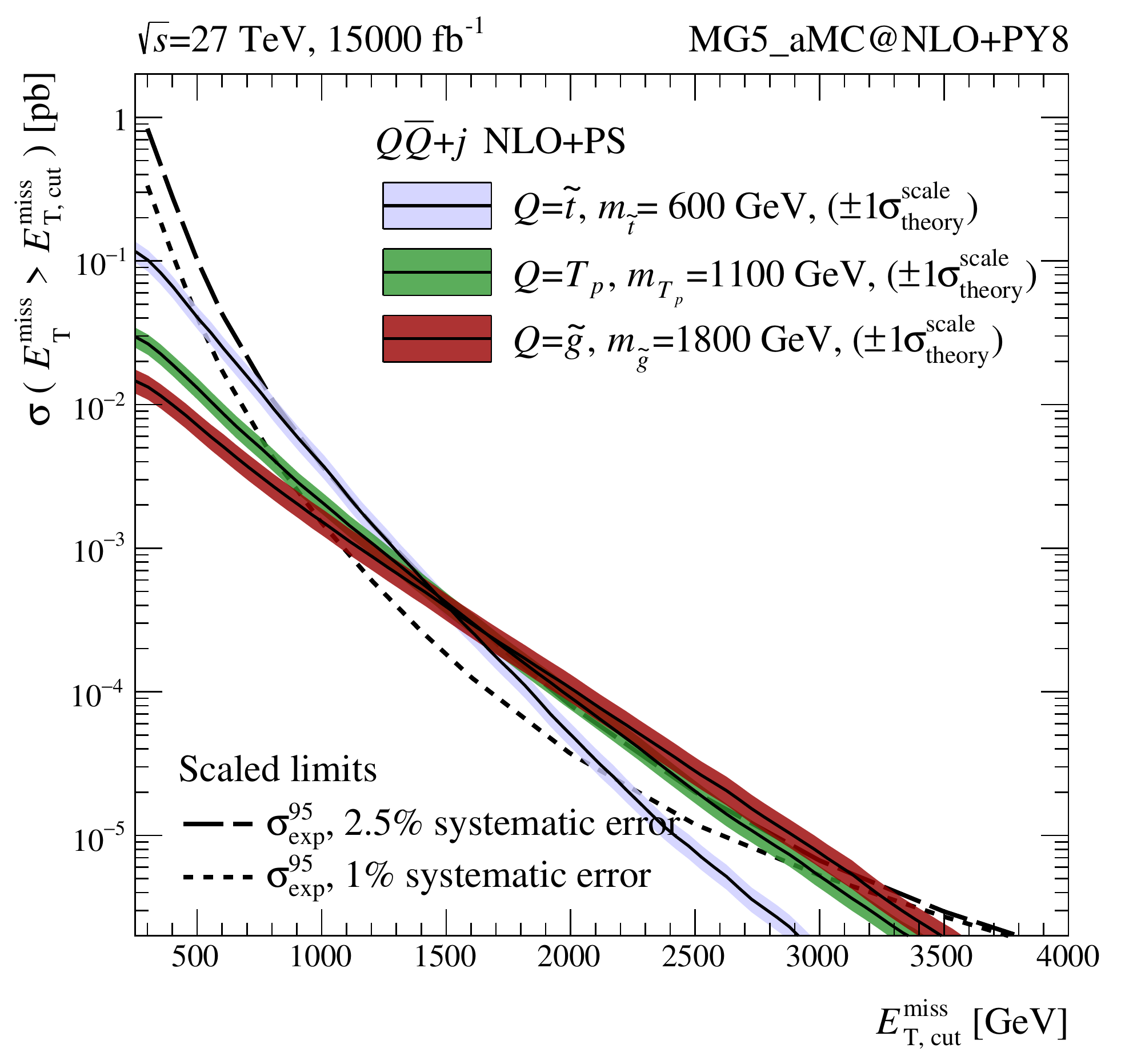} \caption{} 
  \end{subfigure}
     \caption{Same as \figref{fig:comp13TeV} but scaled limits for $\sqrt{s}=27$ TeV assuming (a) $\mathcal{L}=3\iab$ and (b) $15\iab$.}
\label{fig:comp27TeV}
\end{figure*}

\subsection{Properties Determination of Heavy Colored Particles}\label{sec:ident}

We now turn to the possibility of extracting properties of the heavy colored particle $\Q$ from jet behavior within the monojet signature.
As briefly discussed in the introduction, 
asserting color representation and spin of $\Q$ is required to infer information on its mass from cross section measurements (or limits).
Consequently, a single cross section measurement of a particular monojet signal region does not help much in determining the nature of $\Q$.
For example: in \figref{fig:xsec14}, one sees that the production cross section for the process $p p \to \Q \Qbar + j$ with $\ptgen =600$~GeV,
for a top squark $\tilde t$ with  $m_{\tilde t} = $ 400 GeV, a gluino $\tilde g$ with $m_{\tilde g} = $ 900 GeV,
and a  fermionic top partner $\tp$ with $m_{\tp} = $ 600 GeV are roughly within $\Delta\sigma\sim$ 5 - 10 fb of one another.
However, despite this ambiguity, it is still possible to extract information from  the
$pp\to\QQbar + j$ cross section as a function of the leading jet $p_T$, 
which can be measured directly, since it obeys a distinguishing pattern for each $\Q$ hypothesis.
That the nature of $\Q$ is, in part, encoded in this observable reflects a nontrivial interplay 
between $\Q$'s mass, $m_\Q$, its color representation and spin, and the dimensionless ratio $(m_\Q/\ptj)$.
This interplay is what we now discuss.

The first discerning observation is that the $pp\to\QQbar+j$ cross sections do not depend on $\ptgen$ in a universal manner.
Keeping to \figref{fig:xsec14}, one sees that while $\sigma(\QQbar+j; \ptgen =600\GeV)$ are the same at $\sqrt{s}=14$ TeV 
for the $(\Q,m_{\Q})$ configurations under consideration, 
the relative size of $\sigma(\QQbar+j)$ changes with $\ptgen$.
In other words, while $\sigma(\QQbar+j; \ptgen)$ follows an anticipated power-law of $\sigma(\ptgen)\sim (\ptgen)^{-\beta}$,
with $\beta>0$, the precise value of the exponent is dependent on the color and spin structure of $\Q$.
In a particular extreme, the gluino rate is the smallest (largest) of the configurations for $\ptgen$ smaller (larger) than $\ptgen =600\GeV$,
suggesting a smaller $\beta$ than for other $\Q$.
Information on $\Q$ can be extracted from $\sigma(\ptgen)$ by considering its ratio with respect to a benchmark 
$\sigma(\QQbar+j;\ptgen)$.
For example: for the benchmark process $\stp\stpbar+j$ with $m_{\stp}=400\GeV$, 
the $\sqrt s = $ 14 TeV cross section ratios at $\ptgen$ = 400 and 800~GeV
are $\sigma(\go; m_{\go}=900\GeV)/\sigma(\stp)$ = 0.79 and 1.33, respectively,
and similarly $\sigma(\tp;m_{\tp}=600\GeV)/\sigma(\stp)$ = 0.85 and 1.23.
From this one can determine that the change in the cross section ratios over the range 400~GeV $<\ptgen<$ 800~GeV, 
is $\Delta(\sigma/\sigma_{\rm Ref.})/\Delta \ptj\ \sim \mathcal{O}(14\%/100\GeV)$ for gluinos and $\mathcal{O}(9.5\%/100\GeV)$ for fermionic top partners.
Hence, cross section ratios for two different $\Q$ hypotheses is crucially dependent on the choice of $\ptjcut$.

In addition, one can also think about the physics case where there are multiple heavy particles making up the total cross section consistent with a lighter particle.
Take for example that the ratio of $\sigma(\tp\tpbar+j; m_{\tp}=800\GeV)/\sigma(\tp\tpbar+j;m_{\tp}=600\GeV)$ is 0.21 at $\ptgen= 400\GeV$.
Hence, five copies of $\tp$ with mass $m_{\tp}=800\GeV$ can mimic the cross section of a single $\tp$ with $m_{\tp}=600\GeV$ at $\ptgen=400\GeV.$  
At the $\ptgen=800\GeV$, however, $5\times\sigma(\tp\tpbar+j; m_{\tp}=800\GeV)$ / $\sigma(\tp\tpbar+j, m_{\tp}=600\GeV)= 1.25.$  
The one- and five-copy scenarios then predict~
$\Delta(\sigma/\sigma_{\rm Ref.})/\Delta \ptj $ = $ \mathcal{O}(1\%/100\GeV)$ and 
$\Delta(\sigma/\sigma_{\rm Ref.})/\Delta \ptj $ = $ \mathcal{O}(6\%/100\GeV)$, respectively,
thus providing a means to check this potential degeneracy.
Likewise, four copies of $\tp$ with $m_{\tp}=800\GeV$ can mimic the cross section of $m_{\tp}=600\GeV$ at $\ptgen = 800\GeV.$
At $\ptgen=400\GeV,$ however, one finds that $4\times\sigma(m_{\tp}=800\GeV)/\sigma(m_{\tp}=600\GeV)= 0.84$.
For the one- and four-copy scenarios, this leads to the predictions of  
$\Delta(\sigma/\sigma_{\rm Ref.})/\Delta \ptj = \mathcal{O}(1\%/100\GeV)$ and 
$\Delta(\sigma/\sigma_{\rm Ref.})/\Delta \ptj = \mathcal{O}(4\%/100\GeV)$, respectively.

\begin{figure}
\begin{center}
\includegraphics[width=0.5\textwidth]{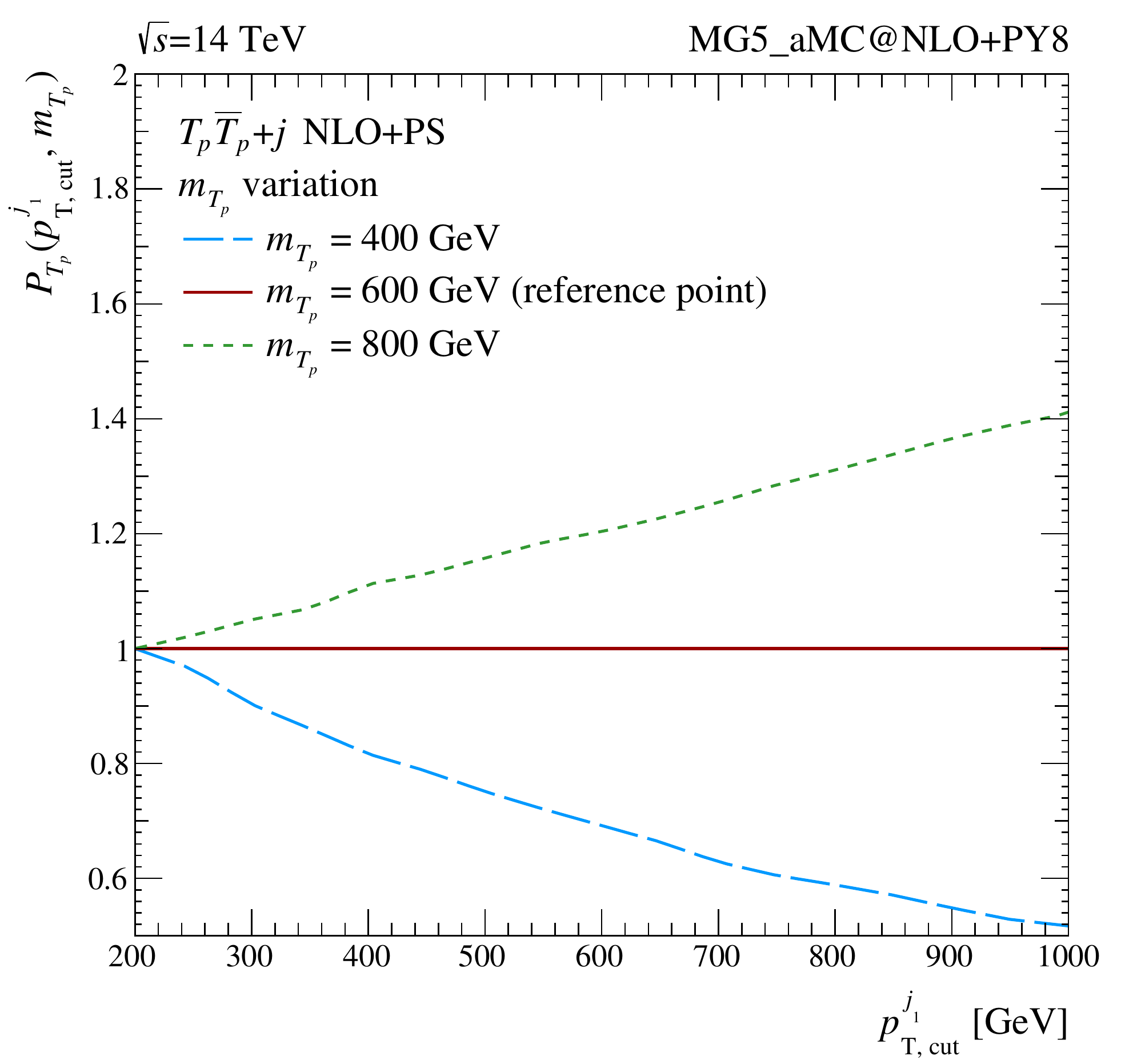}
\end{center}
\caption{
The double cross section ratio $P_{\tp}(\ptjcut,m_{\tp})$, as defined in \eqref{eq:doubleRatMass}, as a function of $\ptjcut$
for fermionic top partners with $m_{\tp} = 400$ and $800\GeV$, and normalization set at $(m_*,p_T^*)=(600\GeV,200\GeV)$.
 } 
\label{fig:xsecRatio1}
\end{figure}

The different dependence on $\ptgen$ observed for gluinos, fermionic top partners, and stops arise from the fact  that heavier particles give rise to harder, 
i.e. less steeply falling, $p_T^{j_1}$ distributions.
The benchmark cases having significantly  different input masses, with $m_{\go}=900\GeV$, $m_{\tp} = 600$ and $800\GeV$, and $m_{\stp}=400\GeV$,
and makes a significant impact on the $\ptj$ dependence.
To isolate this behavior, in \figref{fig:xsecRatio1} we plot the cross section double ratio as a function of $\ptjcut$, 
\begin{eqnarray}
 P_{\Q}(\ptjcut,m_{\Q}) &=&
 \cfrac{\sigma(\QQbar+j; m_{\Q}, \ptjcut)       / 
        \sigma(\QQbar+j; m_{\Q}, \ptjcut=p_{T}^{*})}
       {\sigma(\QQbar+j; m_{\Q}=m_{*}, \ptjcut) / 
        \sigma(\QQbar+j; m_{\Q}=m_{*}, \ptjcut=p_{T}^{*})}.
        \label{eq:doubleRatMass}
\end{eqnarray}
In the top (bottom) single ratio, both cross sections are with respect to the mass $m_{\Q}~(m_*)$ but different $\ptjcut$.
This has the effect of canceling overall color and kinematic factors while isolating logarithmic terms of the form $\log(m_\Q/\ptjcut)^2$.
The double ratio, then, is a measure of this logarithmic dependence with respect to a baseline mass $m_*$ and minimum transverse momentum $p_T^*$. 
For $\Q=\tp$, we choose $m_*=600\GeV$ and $p_T^*=200\GeV$, and plot $P_{\Q}(\ptjcut)$ for $m_{\tp}=400$ and $800\GeV$.
Quantitatively, one sees that the double ratio increases (decreases) by about 50\% at $\ptjcut=800$~GeV for $m_{\tp} =800~(400)$~GeV.
This feature is universal for particles in the same color representation and follows from the nature of massless gauge boson emission in scattering processes.

To better understand this behavior, consider the $pp\to \QQbar+j$ process. 
The $t$-channel propagators gives rise, after phase space integration, to the aforementioned logarithms $\log(m_\Q/\ptjcut)^2$.
In the context of parton shower resummation, this dependence can be interpreted as the likelihood of emitting 
an additional QCD parton with transverse momentum $\ptjcut$, i.e., 
the differential probability is proportional to $d\mathcal{P} = (1/\sigma) d\sigma \propto  \alpha_s(m_\Q)\log(m_{\Q}^2/p_T^{j 2})$.
Hence, a fixed probability $\Delta \mathcal{P}$ implies a fixed $(m_{\Q}/p_T^{j_1})^2$ ratio,
and indicates that increasing $m_{\Q}$ results in $p_T^j$ increasing commensurately. 
Qualitatively, the emission of higher-$p_T$ QCD partons becomes easier for heavier $\Q$ because high-$p_T$ emissions become relatively
soft as $m_{\Q}$ increases. 
This results in a rightward shift of the so-called Sudakov shoulder~\cite{Collins:1984kg,Catani:1997xc}.
For TeV-scale particles, the rightward shift of what constitutes ``soft'' is known to be large; see for example Refs.~\cite{Ruiz:2015zca,Degrande:2016aje}.

The discrimination power of cross section ratios extends if one considers the additional dependency on a collider's beam energy.
In particular, we find that the cross section ratios shift from unity for $\ptgen =600$~GeV at $\sqrt{s}=14$ TeV,
to $\sigma(\go)/\sigma(\stp) = 2.1$  and $\sigma(\tp) /\sigma(\stp )=$ 1.35  at 27 TeV.
Hence, a measurement of the ratio of the cross section at the $\sqrt{s}=14$ and 27 TeV with 30\% accuracy 
can resolve our benchmark gluino, fermionic top parter, and stop  scenario.
Moreover, increasing the beam energy can also significantly improve the signal-over-background ratio, 
thereby enabling measurements of the $\QQbar+j$ cross section for different minimum $p_T^j$ over a wide range of $p_T$.
Measuring the signal cross  section for several $\ptjcut$ at $\sqrt{s}=27\TeV$ with {10\%} accuracy allows 
one to distinguish the benchmark scenario by narrowing down the mass of $\Q$ without $\sqrt{s}= 14\TeV$ information.
In light of this, it is necessary to emphasize that theoretical predictions on total and differential cross sections, as well as their ratios, 
must have the requisite accuracy to make these measurements.

\begin{figure}
\begin{center}
\includegraphics[width=.7\textwidth]{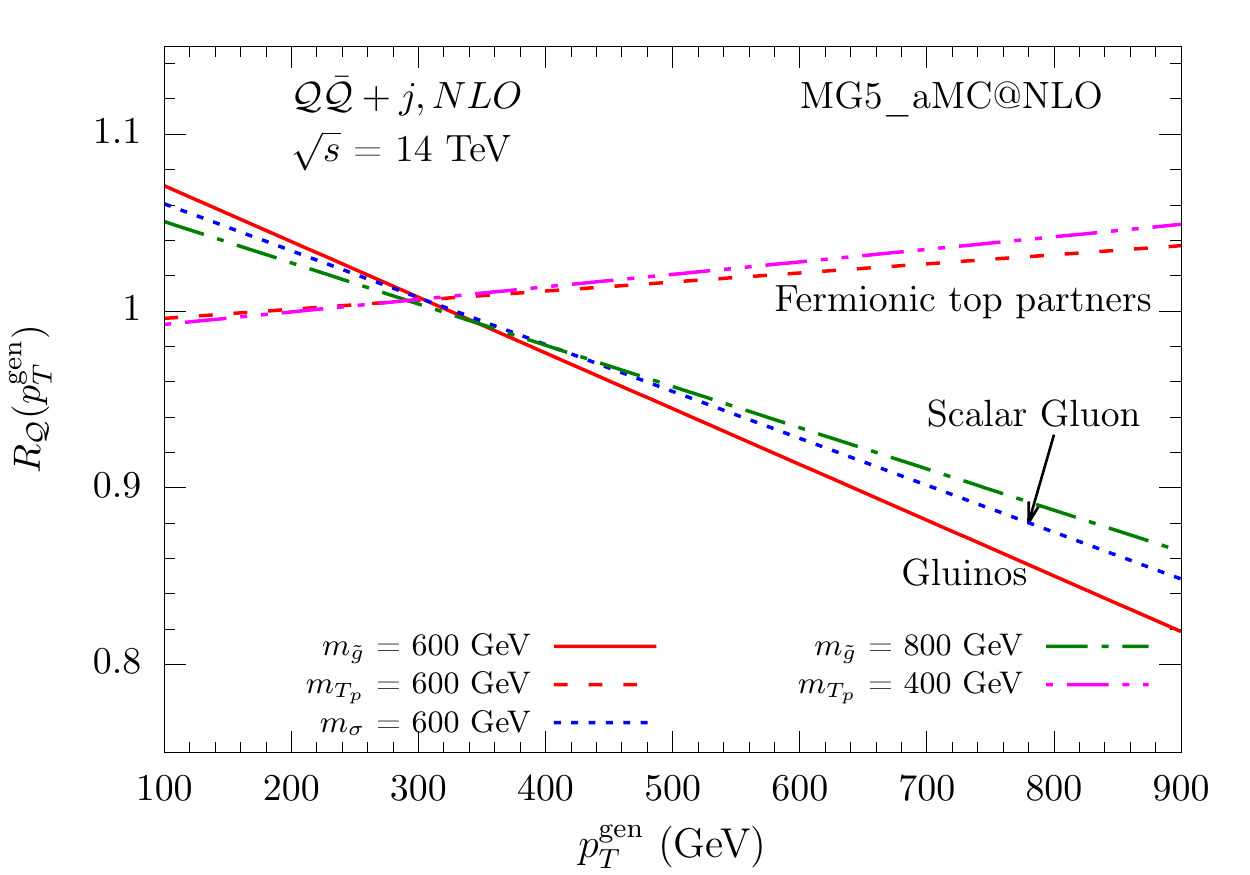}
\caption{The cross section ratio $\sigma( p p \rightarrow \Q \Qbar+j)/\sigma( p p \rightarrow \ststOneJet)$, 
both at NLO in QCD, as a function of a common $\ptgen$, assuming $m_\Q= m_{\stp}$ at $\sqrt{s}=14$ TeV. 
The ratio is normalized to $R(\ptgen)=1$ at $\ptgen=300$~GeV.  
}
\label{fig:xsecRatio2}
\end{center}
\end{figure}

Lastly, we note that the slope of $\sigma(\ptj)$ with respect to $\ptj$ does not significantly depend on either the color structure or spin of $\Q$.
To show this, we display in \figref{fig:xsecRatio2} a second cross section double ratio, \hfill
\begin{eqnarray}
 R_{\Q}(\ptgen,m_{\Q}) &=&
 \cfrac{\sigma(\QQbar+j; m_{\Q}, \ptgen)       / 
        \sigma(\QQbar+j; m_{\Q}, \ptgen=p_{T}^{*})}
       {\sigma(\stp\stpbar+j; m_{\stp}=m_{\Q}, \ptgen) / 
        \sigma(\stp\stpbar+j; m_{\stp}=m_{\Q}, \ptgen=p_{T}^{*})}.
        \label{eq:doubleRatColor}
\end{eqnarray}
The structure of $R_{\Q}(\ptgen,m_{\Q})$ here is analogous to $P_{\Q}(\ptjcut,m_{\Q})$ in \eqref{eq:doubleRatMass},
but differs in that the normalizing process is fixed to $pp\to\stp\stpbar+j$ 
with $\ptgen=p^{*}_{T}=300$~GeV for $\sqrt{s}=14$~TeV, and we vary $pp\to\QQbar+j$ in the upper ratio.
Subsequently, while overall color and kinematic multiplicative factors cancel, 
the relative dependence of $q\overset{(-)}{q}$, $\overset{(-)}{q}g$, and $gg$ scattering within an individual process
does not cancel and is inherently dependent on the color representation. 

From \figref{fig:xsecRatio2}, one can observe that fermionic top partners and stops with same mass have the almost same slope over a wide range of $p_T$. 
On the other hand, because of color and matrix element effects from gluinos, 
the ratio $R_{\tilde{g}}({p^{gen}_T})$ tend to decrease with $p_T$ but again independent to the mass of the heavy particle.
For validation, we consider the case for scalar gluons, known as sgluons \cite{Kribs:2007ac,Plehn:2008ae,GoncalvesNetto:2012nt}, 
and overlay sgluon behavior on the same figure\footnote{The realization of sgluons with a compressed mass spectrum is beyond the scope of this paper.}.
As the plot suggests, the distribution of the hard radiation associated with the colored particle production follows universally
for a given color representation of $\Q$.


\section{Theoretical Uncertainties of the Monojet Process}\label{sec:theoUncertainty}

As investigated in \sectionref{sec:ident}, were one to discover new colored particles at the LHC, or a potential successor experiment,
a measurement of the $\ppQQbarj$ cross section as a function 
of the leading jet's minimum transverse momentum $(\ptjcut)$ can help establish the quantum numbers of $\Q$.
Ascertaining such information, however, requires accurate BSM signal predictions.
For the benchmark scenarios listed in \tableref{tab:model}, 
we find one needs theoretical uncertainties no larger than \confirm{$\mathcal{O}(30\%)$} and \confirm{$\mathcal{O}(5\%)$}, respectively,
on the total inclusive cross section normalizations
and 
on the change of the cross section for $\sigma(\ppQQbarj)$ as a function of $\ptjcut$ per \SI{100}{\giga\electronvolt}.
In this section, we discuss and quantify theoretical uncertainties associated with the monojet signal process.
We particularly investigate (potential) sources of uncertainties when employing the state-of-the-art MC suites \mgpy~and \sherpa.

To investigate theoretical uncertainties associated with the $\QQbarj$ process, we focus on fermionic top partners $\tp$. 
We choose this benchmark because of wide implementability across different event generators as well as its comparability to $\ttbarj$ production in the SM.
We compare several simulation setups and techniques at different levels of precision within QCD; see \sectionref{sec:numeric} for details.
As the relevant signal topology in this work is characterized by the presence of hard QCD  radiation recoiling against the $\tptp$ system, 
our primary benchmark observable is the $\pptptpj$ cross section as a function of the 
transverse momentum of the process' leading, i.e., highest $\pt$, jet $(\ptj)$.

\begin{figure*}
  \centering
  \begin{subfigure}{.49\textwidth}
  \includegraphics[width=\textwidth]{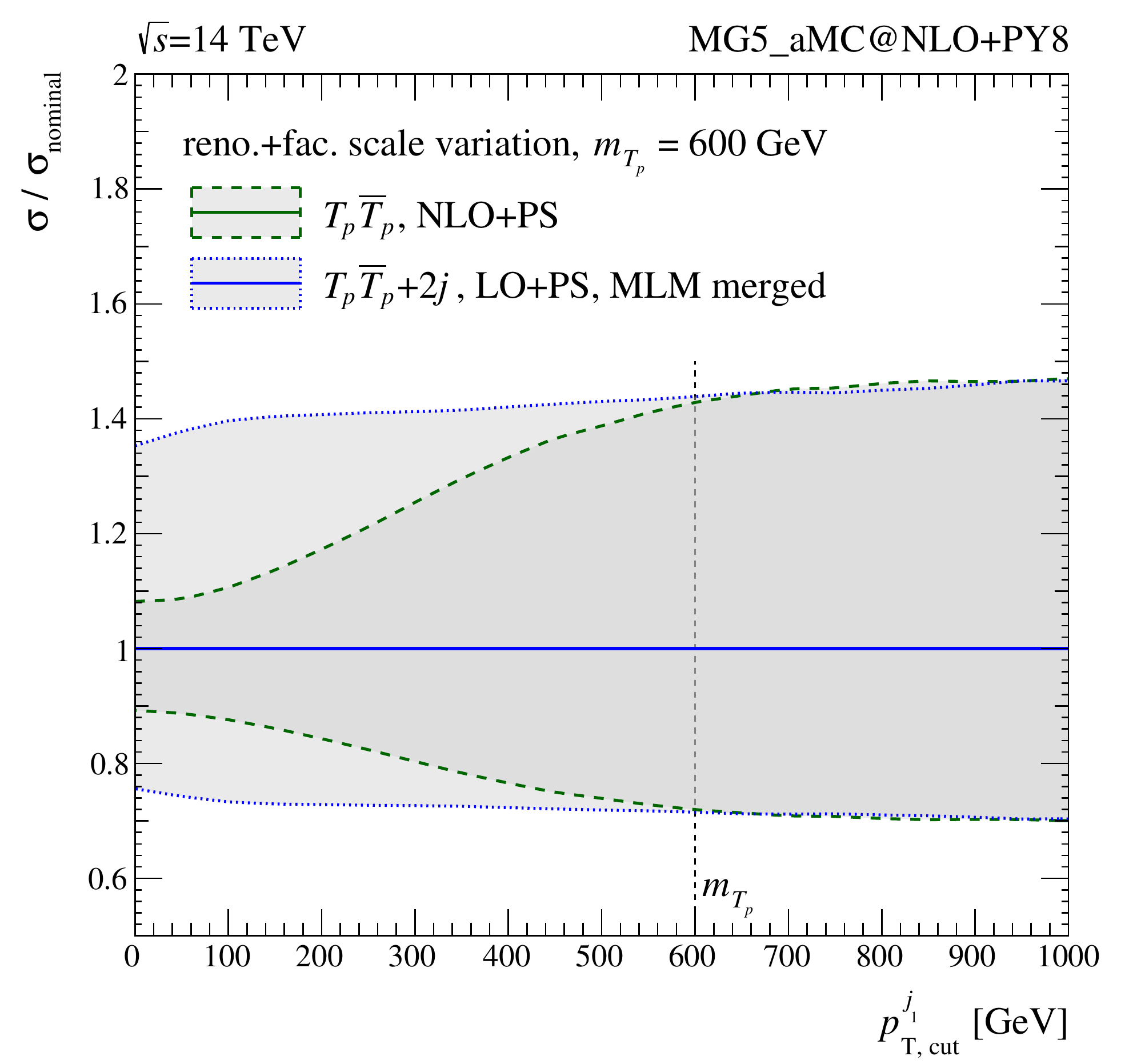} \caption{}	\label{fig:lo-merged-vs-mcnlo}
  \end{subfigure}
  \hfill
  \begin{subfigure}{.49\textwidth}
  \includegraphics[width=\textwidth]{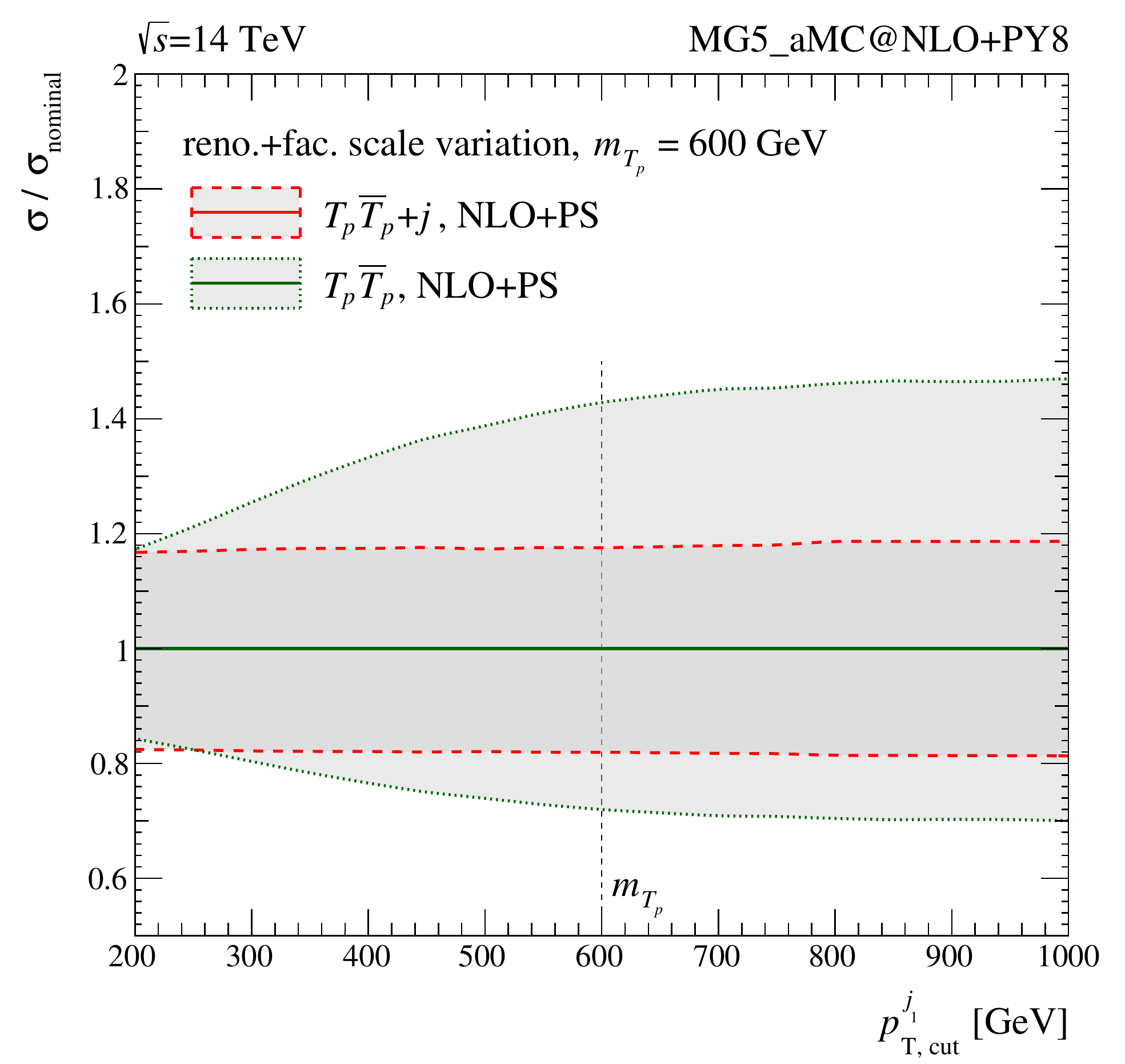} \caption{}	\label{fig:nlo-0j-vs-nlo-1j}
  \end{subfigure}
  \caption{  
  Normalized scale uncertainty bands of the $\pptptpj$ cross section as a function of minimum $\ptj~(\ptjcut)$, for
  (a) the inclusive $\pptptp$ process at NLO+PS (dash) and at LO+PS
  as calculated with up to two additional matrix element-level jets via multijet merging (dot);
  as well as (b) the inclusive $\pptptp$ process at NLO+PS (dot) and the $\pptptpj$ process (no multijet merging) at NLO+PS (dash).  
  }\label{fig:test}
\end{figure*}

We begin with \figref{fig:lo-merged-vs-mcnlo}, where we show the $\pptptpj$ cross section as a function of $\ptj$ 
as derived from the fully inclusive $\pptptp$ calculation at NLO+PS (dash), as obtained from the MC@NLO formalism.
We also show the calculation at LO+PS (dot), as obtained with up to two additional matrix element-level jets via multijet merging.
The curves are shown with their factorization $(Q_F)$ and renormalization $(Q_R)$ variation envelopes and are normalized to the respective nominal prediction.
As discussed in \sectionref{sec:numeric}, the leading jet $\pt$ distribution for 
both calculations is only LO accurate for $\pt^\jet$ comparable to the scale of the hard process.
Hence, the scale uncertainties for $\pt^\jet\gtrsim Q_F,Q_R$ are large, asymmetric, equal for the two calculations, 
and span approximately $+45\%$ to $-30\%$.
For smaller $\pt^\jet$, namely $\pt^\jet \lesssim m_{\tp}=600\GeV$, one observes that the two uncertainty envelopes begin to differ.
Whereas the uncertainty for the multijet calculation reduce only slightly for decreasing $\pt^\jet$,
the MC@NLO uncertainty reduces to roughly the $\pm10\%$ level.
The difference originates from virtual corrections present in the NLO+PS calculation,
which soften dependencies on $Q_F,Q_R$, but are obviously absent in the LO+PS calculation.
The Sudakov-like factor in the multijet merging prescription only partially reduces the dependence on $Q_F$ 
by matching low-$\pt$ QCD emissions in the hard matrix element with those in the PDF.
Hence, even for observables that are formally of the same precision,
the presence of all $\mathcal{O}(\alpha_s)$ terms in the NLO+PS calculation leads to a smaller scale dependence at low $\ptj$
than in the merged LO+PS calculation.

The small scale uncertainty observed for the lowest $\ptj$ 
suggests high theoretical precision is achievable with NLO+PS computations obtained via the MC@NLO prescription.
However, unlike pure FO calculations, calculations matched to parton showers 
possess the additional dependence on the parton shower starting scale $Q_S$.
Within the MC@NLO formalism, $Q_S$ controls whether the leading $\mathcal{O}(\alpha_s)$ emission
beyond the Born process is included in the FO matrix element or the all-orders parton shower.
Loosely speaking, QCD radiations with $\pt^\jet$ above (below) $Q_S$ originate from the hard matrix element (parton shower).
As pointed out in Refs.~\cite{Hoeche:2011fd,Jones:2017giv}, 
lowest order-accurate observables, e.g., the $\ptj$ distribution when the $\pptptp$ process is evaluated at NLO+PS,
and processes that possess large virtual corrections suffer from ambiguities when choosing $Q_S$.
This manifests as a strong dependence on $Q_S$, and hence a large deviation.

We assess this behavior in \figref{fig:mus-var} by plotting the $\pptptpj$ cross section, 
derived from the inclusive $\pptptp$ calculation at NLO+PS, as a function of~$\ptjcut$. 
We assume multiplicative variations of the default parton shower scale, given in \eqref{eq:showerScaleUnc},
but fix the factorization and renormalization scales to their central values.
Rates are normalized to the fixed order NLO (fNLO) prediction.
The $Q_S$ variation of the NLO+PS result with respect to the default $Q_S$ choice spans roughly $\pm25\%$, 
with the dependence increasing (decreasing) for larger (smaller) values of $Q_S$ over the range of $\ptj$ considered.
The shower scale variation amounts to absolute deviations from the fNLO result up to about $+60\%$ and $-35\%$.	
For vanishing $\pt^\jet$, one should take caution in interpreting the vanishing shower scale uncertainty.
In this limit, the FO calculation is unphysical. 
The FO calculation possess an integrable singularity at $\pt^\jet=0$ that leads to arbitrarily large cross sections. 
As a result, the ratio of the NLO+PS-level cross section, a finite and physical quantity, to the fNLO cross section, the unphysical quantity, vanishes as $\pt^\jet/m_{\tptp}\to0$.
For the curves with shower scale multiplier 0.5 and 1, we observe the curves are flat beyond $\sim 300$ GeV and 600 GeV, respectively, as expected. However, the curves with shower scale multiplier larger than 1, we observe a crossing at $p_T \sim 700$ GeV. The large shower starting scale introduces high $p_T$ initial state radiations, which is neither soft nor collinear to the incoming parton. If the high $p_T$ jets generated from the parton shower of Born-type $S$ events immersed in this region, then the cross section is not reliable. Therefore, the curves (and the crossing) are not trustworthy beyond the top partner mass, or equivalently $\sqrt{\hat{s}}/2$. A better way to estimate the $pp\rightarrow T_p \bar{T}_p + j$ cross section (and associated shower scale uncertainty) would be to consider the process $pp\rightarrow T_p \bar{T}_p+j$ at NLO+PS.

In light of the large theoretical uncertainties in the $\ppQQbarj$ cross section stemming from $Q_F,~Q_R,$ and $Q_S$, 
it is clear that the precision achieved with the aforementioned methods is insufficient for distinguishing $\Q$ candidates.
To explore if such precision is yet still possible with presently available general-purpose Monte Carlo technology,
we consider the cross section obtained from the $\pptptpj$ process itself at NLO+PS.
In \figref{fig:nlo-0j-vs-nlo-1j} we plot the normalized $\pptptpj$ cross section with its $Q_F,~Q_R$ envelope again as a function of $\ptjcut$,
but as derived from the $\pptptp$ (dot) and $\pptptpj$ (dash) calculations at NLO+PS.
We observe that the uncertainty reduces to a largely uniform band that is \confirm{just shy of $\pm20\%$} for the $\tptpj$ calculation.
Both uncertainties are comparable for $ \ptj \ll m_{\tp}$;
for $\ptj \gg m_{\tp}$, however, the $\tptpj$ uncertainty is about about \confirm{$40-60\%$} smaller.

\begin{figure}
  \centering
  \includegraphics[width=.49\textwidth]{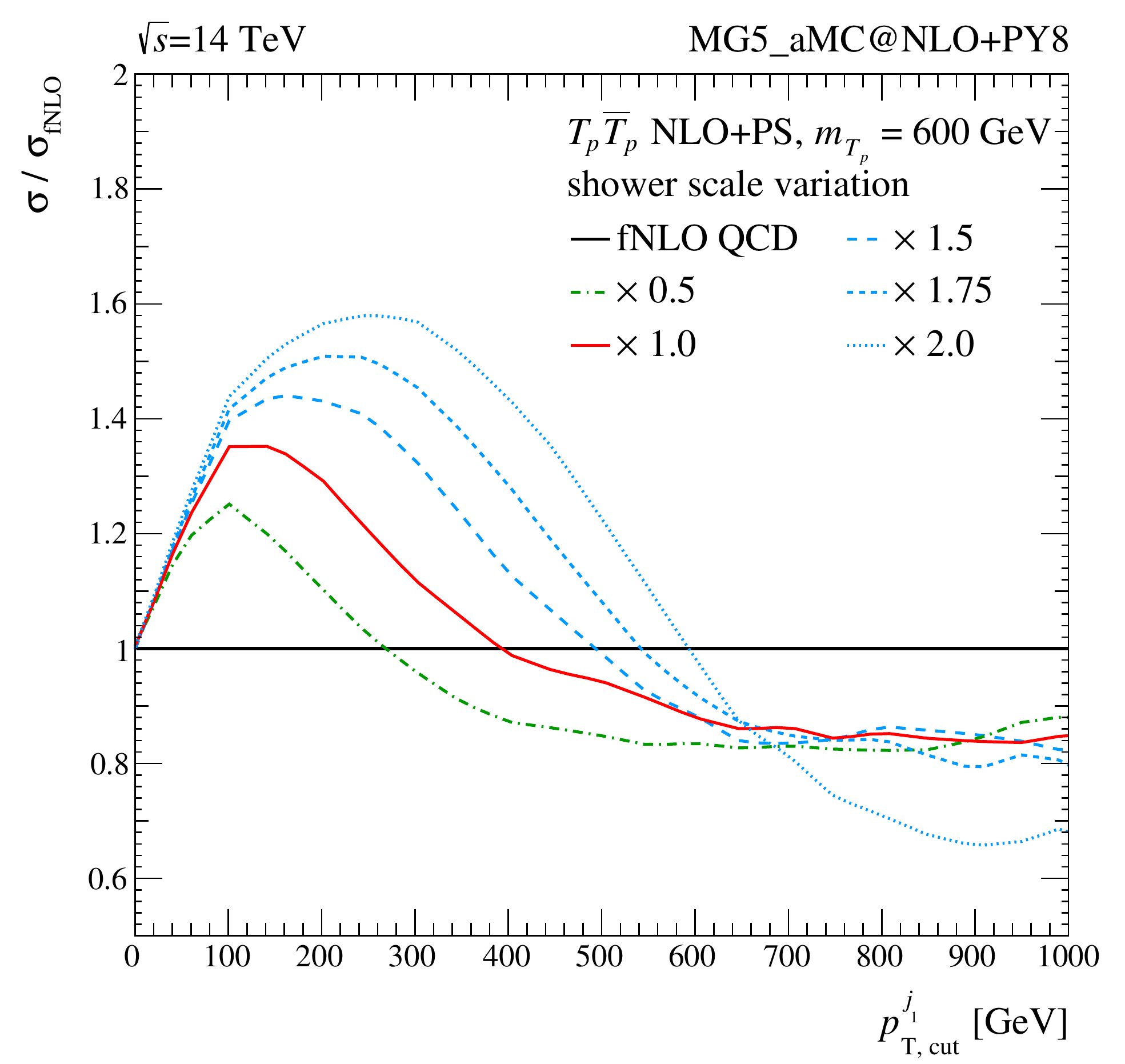}
  \caption{
  The $\pptptpj$ cross section, derived from the inclusive $\pptptp$ calculation at NLO+PS,
  as a function of $\ptjcut$, assuming multiplicative variations of the default parton shower scale.
  Rates are normalized with respect to the fNLO prediction.
    }
\label{fig:mus-var} 
\end{figure}

It is worth reiterating that the FO $\tptpj$ calculation requires a minimum $\pt^\jet$ selection to regulate matrix element poles and,
since no Sudakov suppression is applied to low $\pt$ QCD emissions in this case, 
to ensure perturbative stability, in the Collins-Soper-Sterman (CSS) sense~\cite{Collins:1984kg}.
We safeguard against the need for $k_T$-resummation on the leading jet
by using the (somewhat conservative) CSS consistency requirement from Ref.~\cite{Degrande:2016aje}.
For $m_{\tp}=600\GeV$, and hence $m_{\tptp}>1.2\TeV$, one needs $\ptjcut\gtrsim 200-250\GeV$
to force collinear logarithms from $t$-channel exchanges, which scale as $\delta\sigma\sim \alpha_s\log(m_{\tptp}/\pt^{j})^2$, 
to be much smaller than $1$.

There is now an apparent conflict between theoretical needs and computational capabilities:
While $\pptptpj$ at NLO+PS provides improved control and stability over $Q_F$ and $Q_R$,
the calculation is only meaningful for sufficiently large $\ptj$.
The $\pptptp$ process at NLO+PS, on the other hand, extends the support for $\ptj$ to low $\pt$ but at the cost of a larger uncertainties.
Such demands, however, are precisely resolved with multijet merging at NLO+PS, i.e., MEPS@NLO.
As the CKKW MEPS@NLO prescription is natively available in the event generator \sherpa,
for the remainder of this section we use \sherpa~to further quantify scale uncertainties in $\tptp$ production.
However, before proceeding, we non-trivially demonstrate that at least up to the QCD order presently being investigated, 
both \mgpy~and \sherpa~ predictions are in agreement with one another.

\begin{figure*}
  \centering
  \begin{subfigure}{.49\linewidth}
  \includegraphics[width=\textwidth]{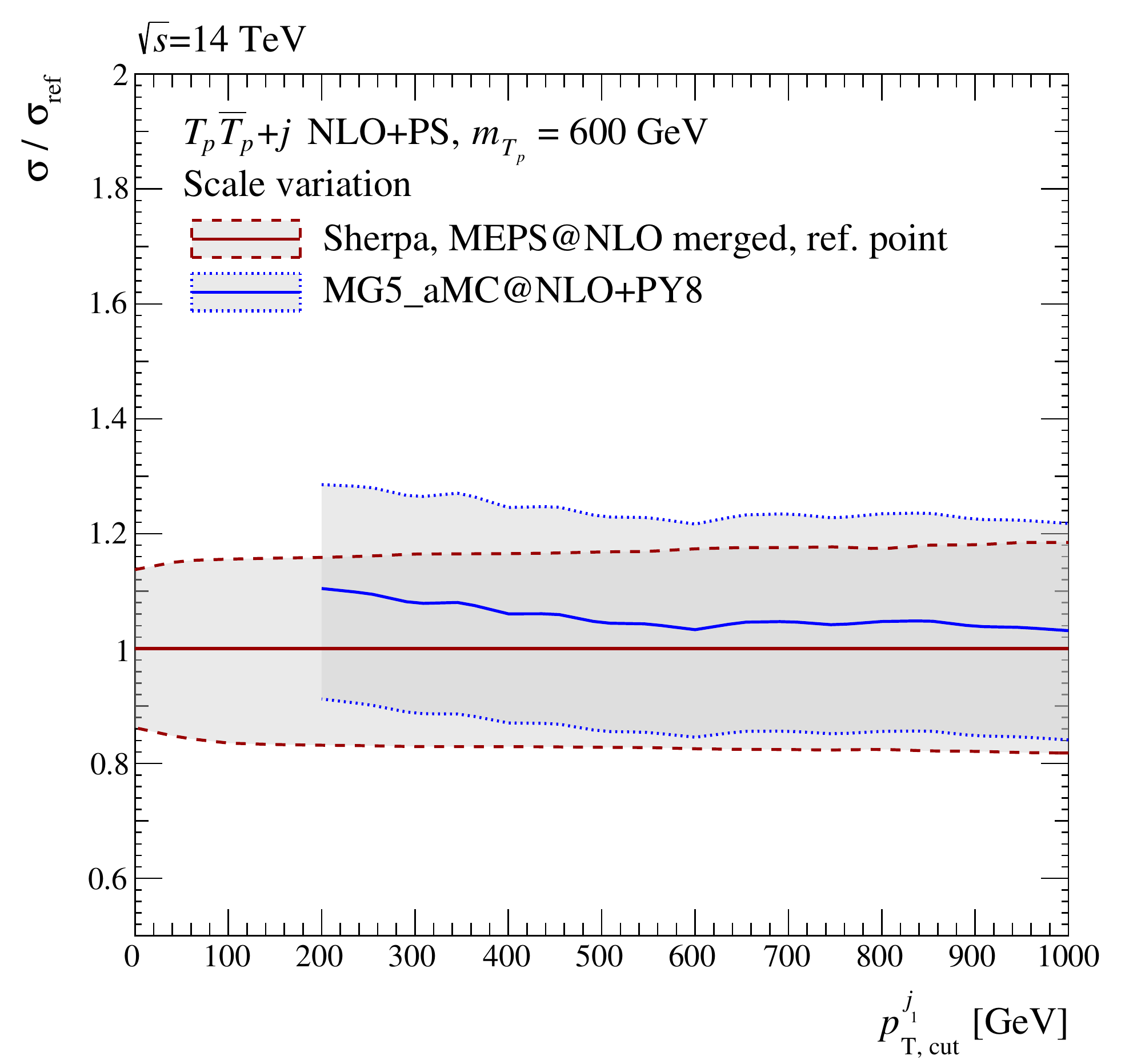}
  \caption{Comparison of the $\pptptpj$ cross section with $Q_F,~Q_R$ uncertainty band
  as derived from the $\pptptpj$ process at NLO+PS (\mgpy)
  and from the $\pptptp$ process with MEPS@NLO multijet merging to one additional jet (\sherpa).
  Curves are individually normalized to their central value.
  }
  \label{fig:shvsmgL}
  \end{subfigure}
  \hfill
  \begin{subfigure}{.49\linewidth}
  \includegraphics[width=\textwidth]{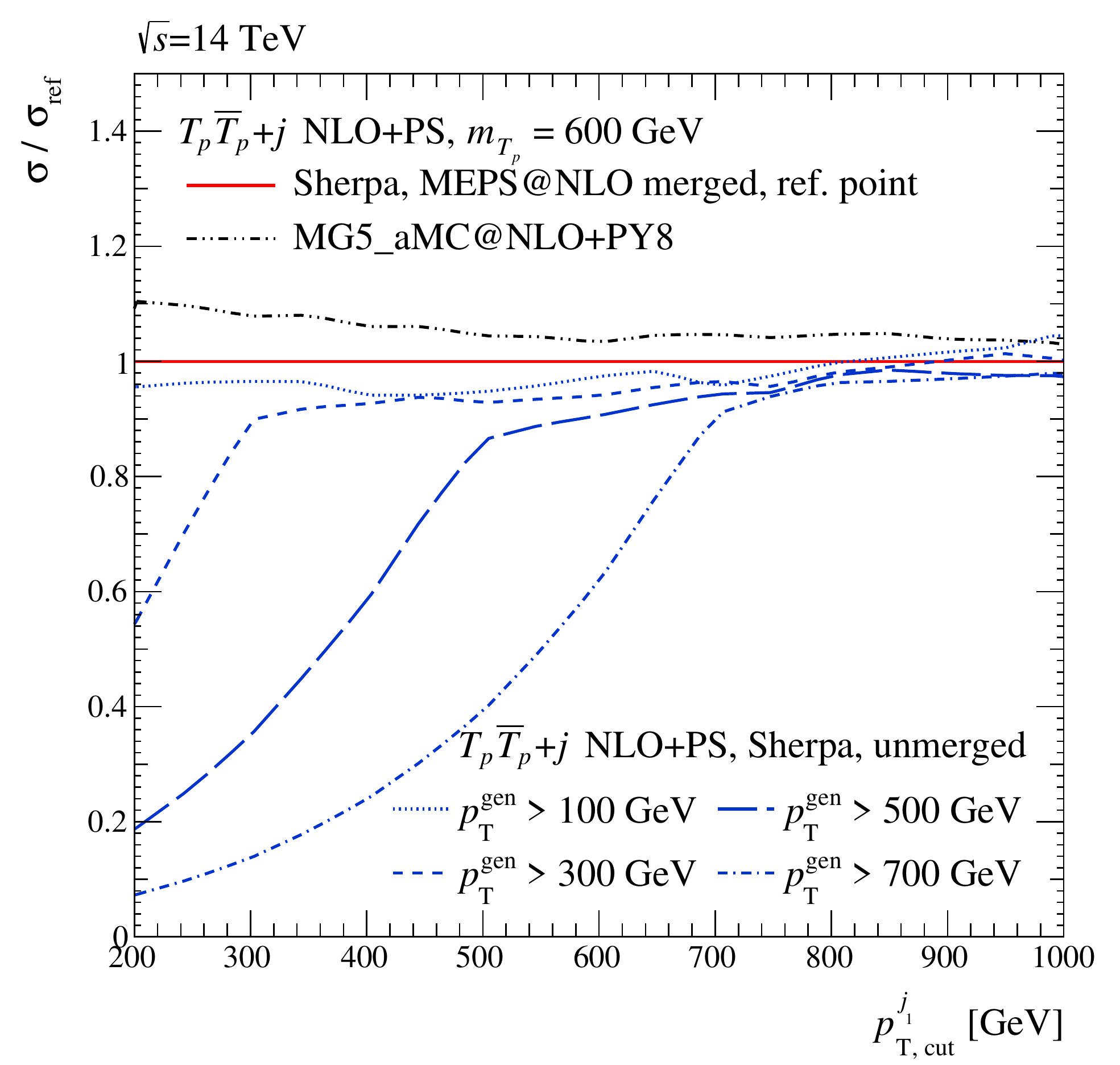}
  \caption{
  The same quantities as (a) but overlaid with the individual $\pptptpj$ at NLO+PS samples (\sherpa) that constitute the MEPS@NLO result. 
  Unmerged NLO+PS calculations are distinguished by their $\ptgen$ selection on $\ptj$.
  \newline \hphantom{0}
  }
  \label{fig:shvsmgR}
  \end{subfigure}
  \caption{ Comparison of the $\pptptpj$ cross section estimated using \mgpy ~and \sherpa~ with scale uncertainty band.}
\end{figure*}

In \figref{fig:shvsmgL}, we plot the $\pptptpj$ cross section and its scale uncertainty band as a function of $\ptjcut$
as derived from the $\pptptpj$ process at NLO+PS using \mgpy,
as well as from the $\pptptp$ process at NLO+PS merged with up to one additional NLO-accurate, matrix element-level jet using \sherpa.
The curves are normalized to the \sherpa~rate.
For $\ptjcut>200\GeV$, one sees that the \sherpa~result, much like the \mgpy~result, 
possesses a stable uncertainty envelope spanning roughly $\pm18\%$.
Moreover, the central value of \sherpa~remains only $3-11\%$ below \mgpy, 
and is consistent with the difference choice of $\alpha_s(Q_s)$ employed by~\sherpa.
For $\ptj<200\GeV$, the MEPS@NLO calculation reveals that the $Q_F$ and $Q_R$ dependence remains smooth, 
and tapers down to $\pm16\%$ as $\ptjcut$ is relaxed.
In comparison to the inclusive $\pptptp$ at NLO+PS in \figref{fig:lo-merged-vs-mcnlo},
the uncertainty band here is $\mathcal{O}(5\%)$ larger in both directions.
The slightly larger uncertainty here is consistent with the presence of additional soft $\mathcal{O}(\alpha_s^2)$ radiation terms
whose scale dependence would otherwise be stabilized by $\mathcal{O}(\alpha_s^2)$ virtual terms.

As described in \sectionref{sec:numeric}, to construct the MEPS@NLO sample, 
various $\pptptpj$ NLO+PS computations at increasing $\ptgen$ are necessary to populate the $\pt^\jet$ tail for $\pt^\jet \gg \ptgen$.
To further demonstrate the comparability of the \mgpy~and~\sherpa~curves,
we show in \figref{fig:shvsmgR} the same quantities plotted in \figref{fig:shvsmgL} for $\ptjcut>200\GeV$
but overlaid with the individual, unmerged $\pptptpj$ samples.
For the range of $\ptjcut$ considered, we find that the difference between the \mgpy~NLO+PS prediction and the \sherpa~NLO+PS predictions for $\ptjcut>\ptgen$ 
differ only by about $0-15\%$ again within the $Q_F,~Q_R$ uncertainty.

\begin{figure*}
  \centering
  \begin{subfigure}{.49\textwidth}
  \includegraphics[width=\textwidth]{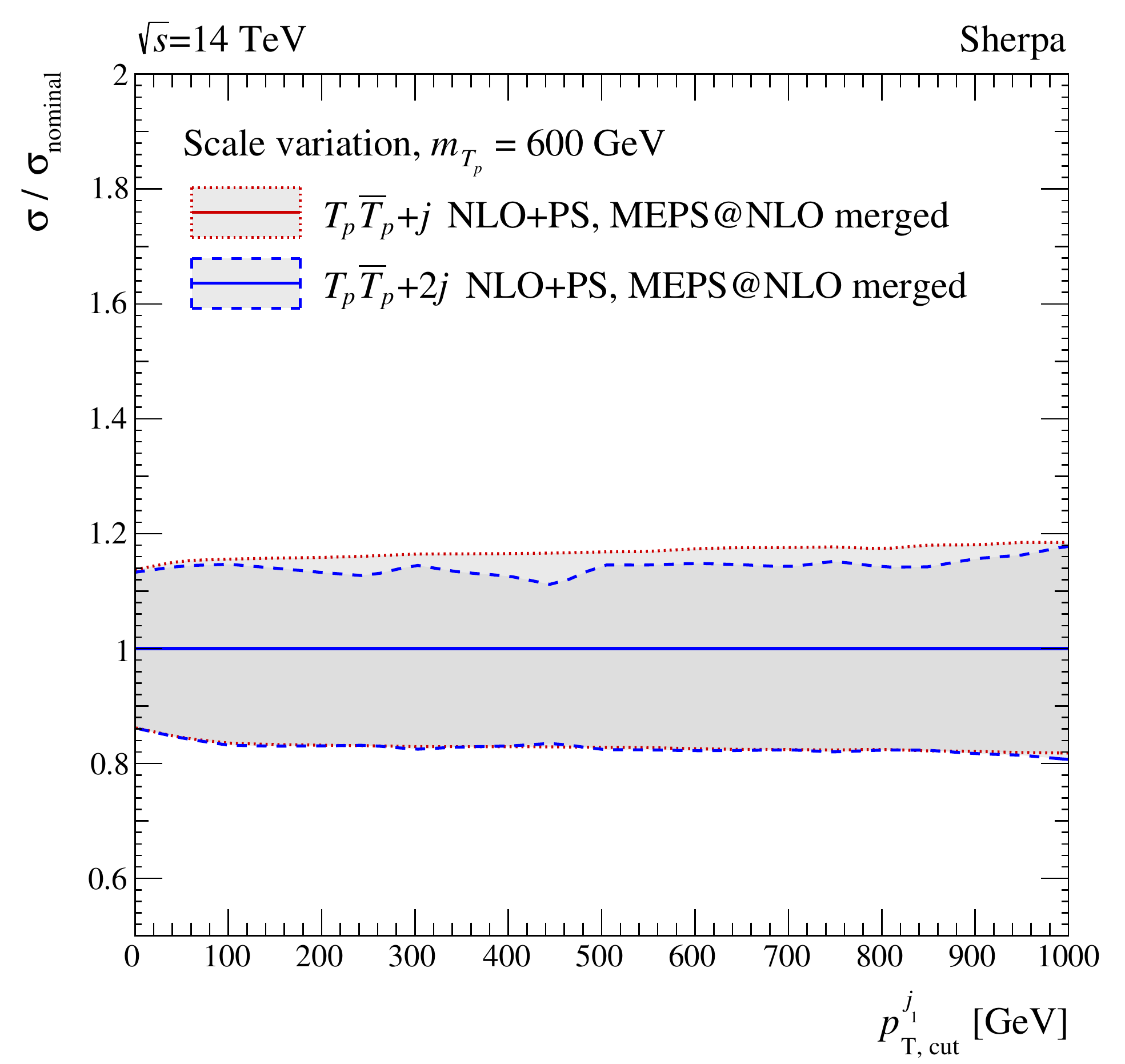} 
  \caption{}\label{fig:meps-muf-mur-ptj1}
  \end{subfigure}
  \hfill
  \begin{subfigure}{.49\textwidth}
  \includegraphics[width=\textwidth]{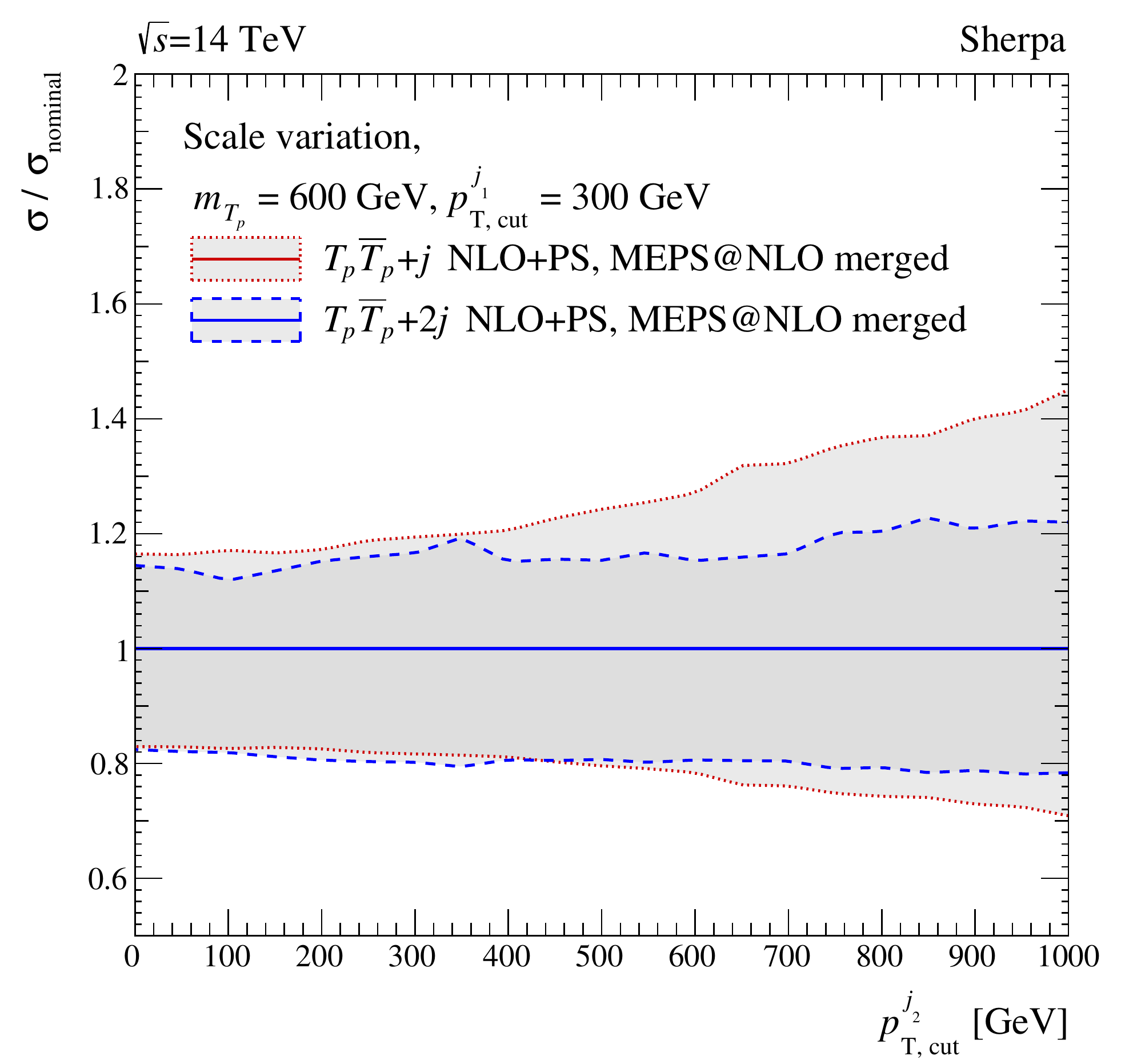} 
  \caption{}\label{fig:meps-muf-mur-ptj2}
  \end{subfigure}
  \caption{Factorization and renormalization scale uncertainty bands for 
  (a) the $\pptptpj$ cross section as a function of $\ptjcut$ and
  (b) the $\pptptpjj$ cross section as a function of $\ptjjcut$ assuming $\ptjcut=300\GeV$,
  for the $\pptptp$ process with MEPS@NLO matching up to one (dark curve) or two (light curve) additional jets.
  Curves are normalized to the $2\jet$ MEPS@NLO rate.
  }
  \label{fig:meps-muf-mur}
\end{figure*}


A theoretical difficulty of the inclusive monojet signature is ensuring perturbative control over predictions,
despite complicated, multi-scale requirements on jet momenta and particle multiplicity.
Hence, it is important to ensure that sub-leading jets in the monojet processes can be modeled as well as the leading jet.
To investigate whether such uncertainties for subleading jets are attainable, we compare MEPS@NLO uncertainties
for the $\pptptp$ process when merged with different jet multiplicities. 
As a control, we show in \figref{fig:meps-muf-mur-ptj1} the (normalized) $\pptptpj$ cross section and uncertainty band as a function of $\ptjcut$
for the $\pptptp$ process at NLO+PS merged 
with up to one (dark curve) or two (light curve) additional NLO-accurate, matrix element-level jets.
No significant difference between the two curves is observed.
One does not expect to see such deviations for this observable as the two calculations are identical at this order of $\alpha_s$.
The difference appear at one order higher: 
At the matrix element level, the two calculations possess up to two and three hard QCD emissions, respectively.
In the $1\jet$ MEPS@NLO calculation, the second jet is LO+LL (alternatively, LO+PS) accurate.
In the $2\jet$ MEPS@NLO calculation, the second jet is NLO+LL accurate and the third jet is LO+LL accurate.
Therefore, in \figref{fig:meps-muf-mur-ptj2}, we plot the $\pptptpjj$ cross section as a function of minimum $\ptjj$ for the two calculations.
In analogy to the comparison of $\tptp$ and $\tptpj$ production at NLO+PS in \figref{fig:nlo-0j-vs-nlo-1j},
we observe here that the $1j$ MEPS@NLO prediction suffers from an uncertainty spanning roughly $+45\%$ to $-30\%$
whereas the $2\jet$ MEPS@NLO rate exhibits a largely uniform uncertainty of about $\pm22\%$.
For vanishing $\ptjjcut$, the uncertainties become equal since the observable is no longer sensitive to such a high order of $\alpha_s$.
With successive MEPS@NLO matching to higher jet multiplicities, one expects a comparable reduction of dependence on  $Q_F,~Q_R$ in the corresponding cross section.
However, one should not expect the uncertainty to drop below the observed 
$\mathcal{O}(10-15\%)$ without first accounting 
for missing two-loop terms at $\mathcal{O}(\alpha_s^2)$.

\begin{figure}
  \begin{subfigure}{.49\textwidth}
  \includegraphics[width=\textwidth]{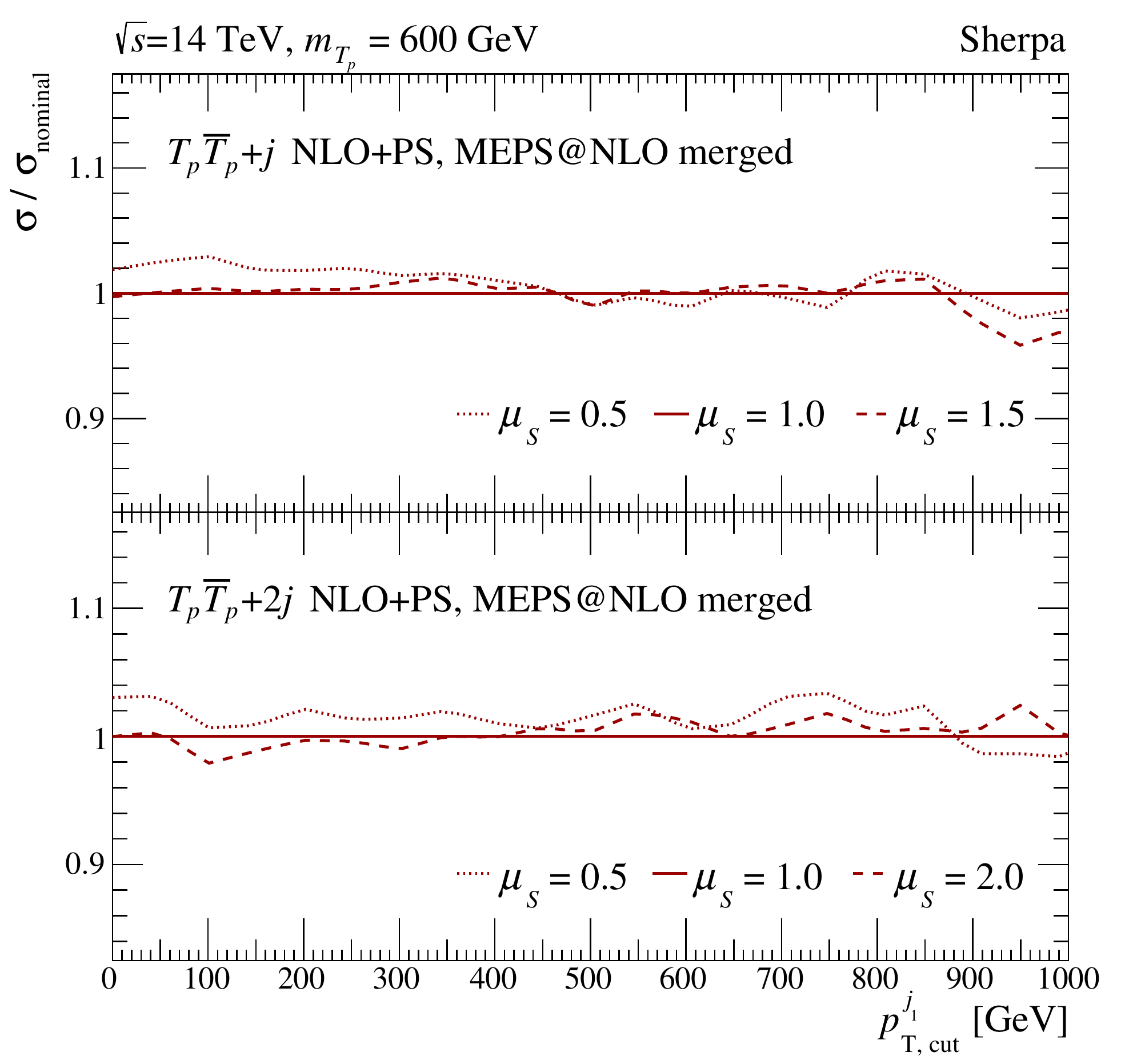}\caption{}\label{fig:meps-muS-ptj1}
  \end{subfigure}
  \hfill
  \begin{subfigure}{.49\textwidth}
  \includegraphics[width=\textwidth]{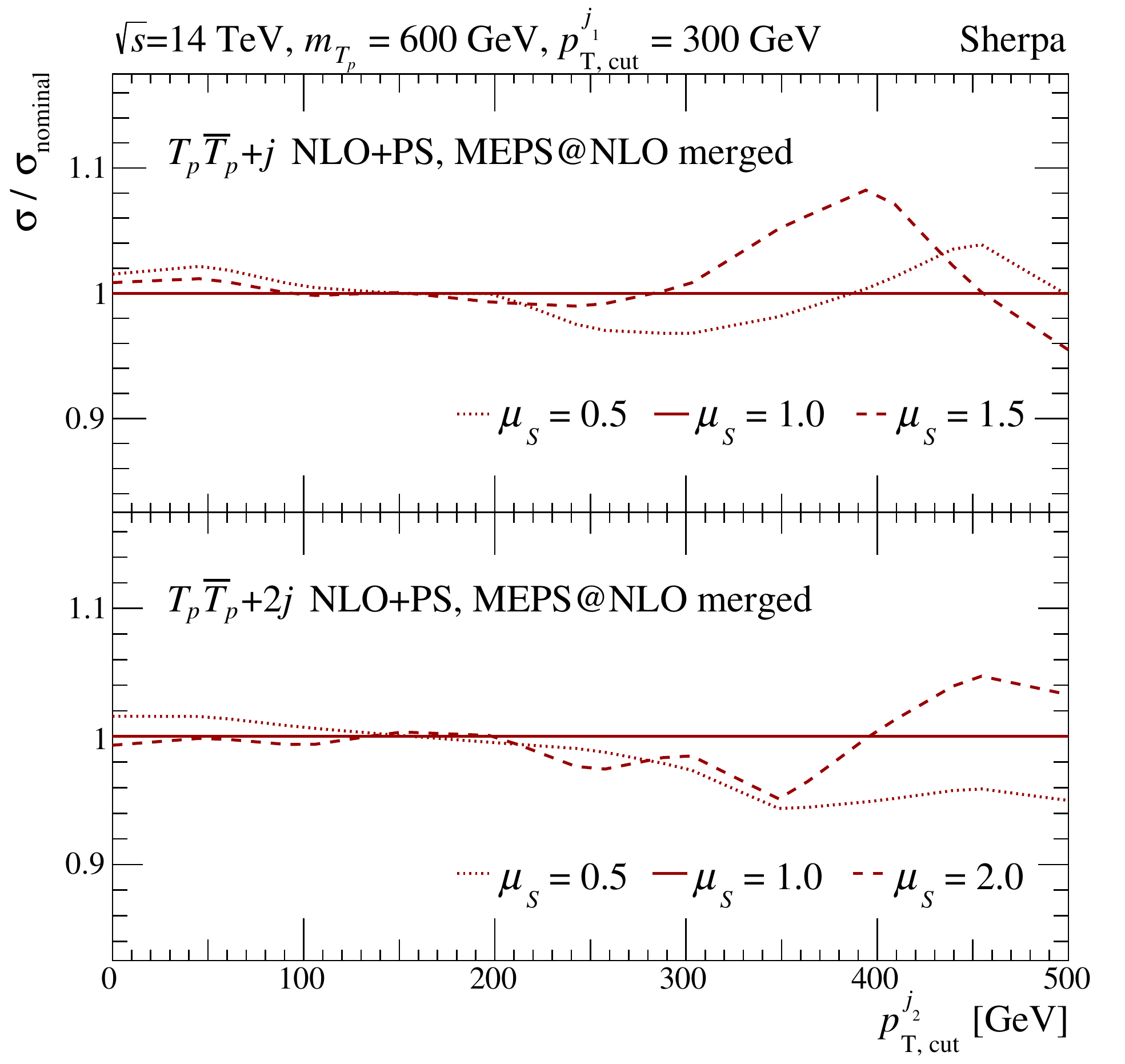}\caption{}\label{fig:meps-muS-ptj2}
  \end{subfigure}
  \caption{Parton shower scale dependence for 
  (a) the $\pptptpj$ cross section as a function of $\ptjcut$ and
  (b) the $\pptptpjj$ cross section as a function of $\ptjjcut$ assuming $\ptjcut=300\GeV$,
  for the $\pptptp$ process with MEPS@NLO matching up to one (upper) or two (lower) additional jets.
  Curves are normalized to the respective rate assuming the default shower scale.
  }
  \label{fig:meps-muS}
\end{figure}

We now turn to the issue of parton shower scale uncertainties for processes evaluated with MEPS@NLO.
Following the same variation procedure as before, we plot in  \figref{fig:meps-muS-ptj1} and  \figref{fig:meps-muS-ptj2}, respectively,
the $Q_S$ dependence for the $\tptpj$ and $\tptpjj$ cross sections as a function of minimum $p_T$ 
for the leading $(p_T^{j_1})$ and subleading $(p_T^{j_2})$ jet.
The upper (lower) panel corresponds to the $1\jet$ ~($2\jet$) MEPS@NLO prediction.
All curves are normalized to the appropriate cross section evaluated at the default scale choice.
We find for both calculations that the uncertainty in the $\tptpj$ cross section is very small, 
reaching maximally to \confirm{$\pm4\%$} over the range of $\ptj$ considered, and comparable to our MC, i.e., statistical, uncertainty.
Relative to the $\tptp$ at NLO+PS uncertainty in \figref{fig:mus-var}, which exhibits deviations up to \confirm{$\pm25\%$},
this represents significant reduction in $Q_S$ dependence.
For the $\tptpjj$ cross section, we see a larger scale dependence in both the $1\jet$ and $2\jet$ calculations.
The uncertainties for the two rates are comparable, reaching about \confirm{$+10$ to $-5\%$}, and again comparable to our MC uncertainty.

\begin{figure*}[!t]
  \centering
  \begin{subfigure}{.49\textwidth}
  \includegraphics[width=\textwidth]{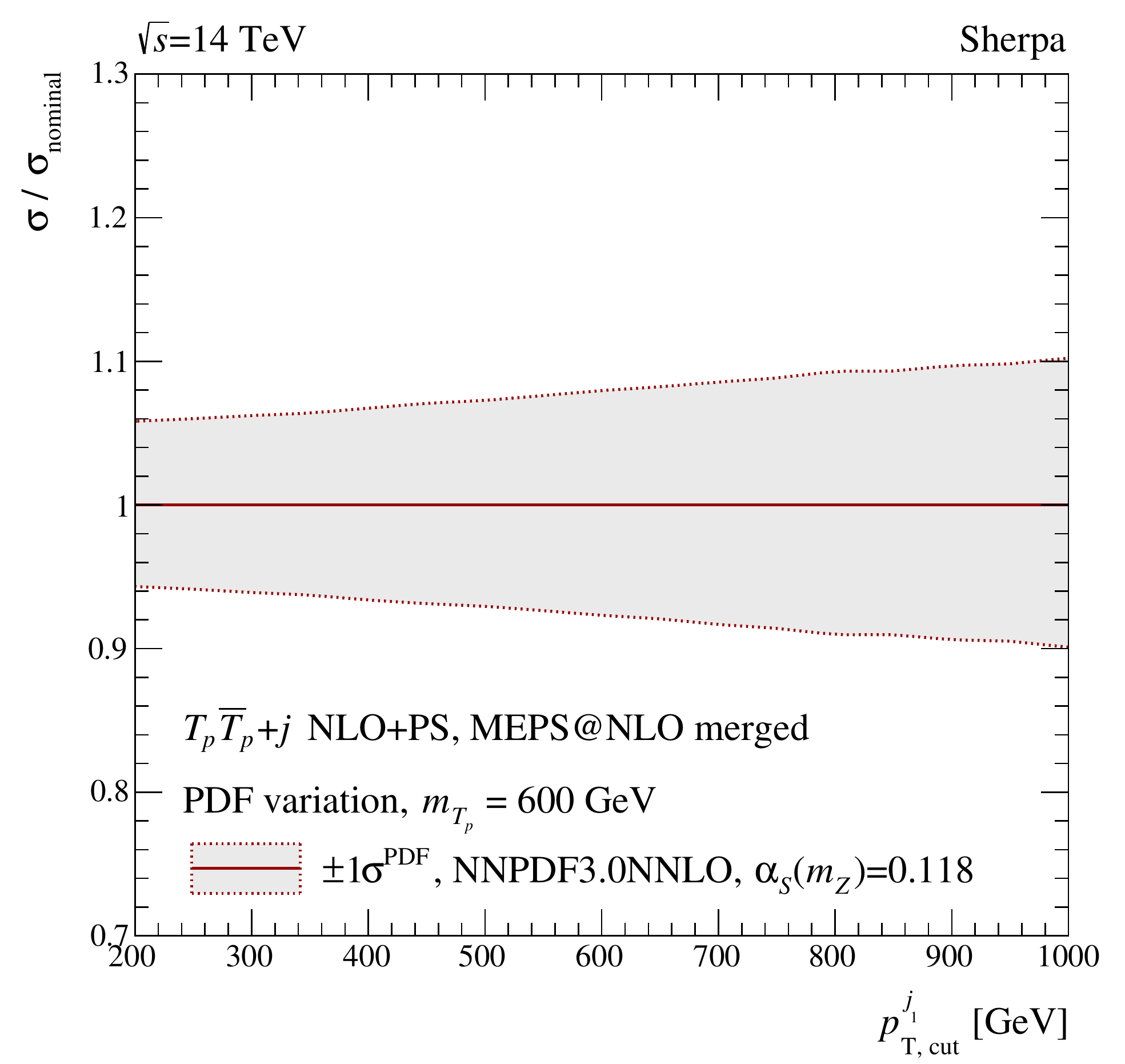}\caption{}\label{fig:meps-PDF-PDF}
  \end{subfigure}
  \hfill
  \begin{subfigure}{.49\textwidth}
  \includegraphics[width=\textwidth]{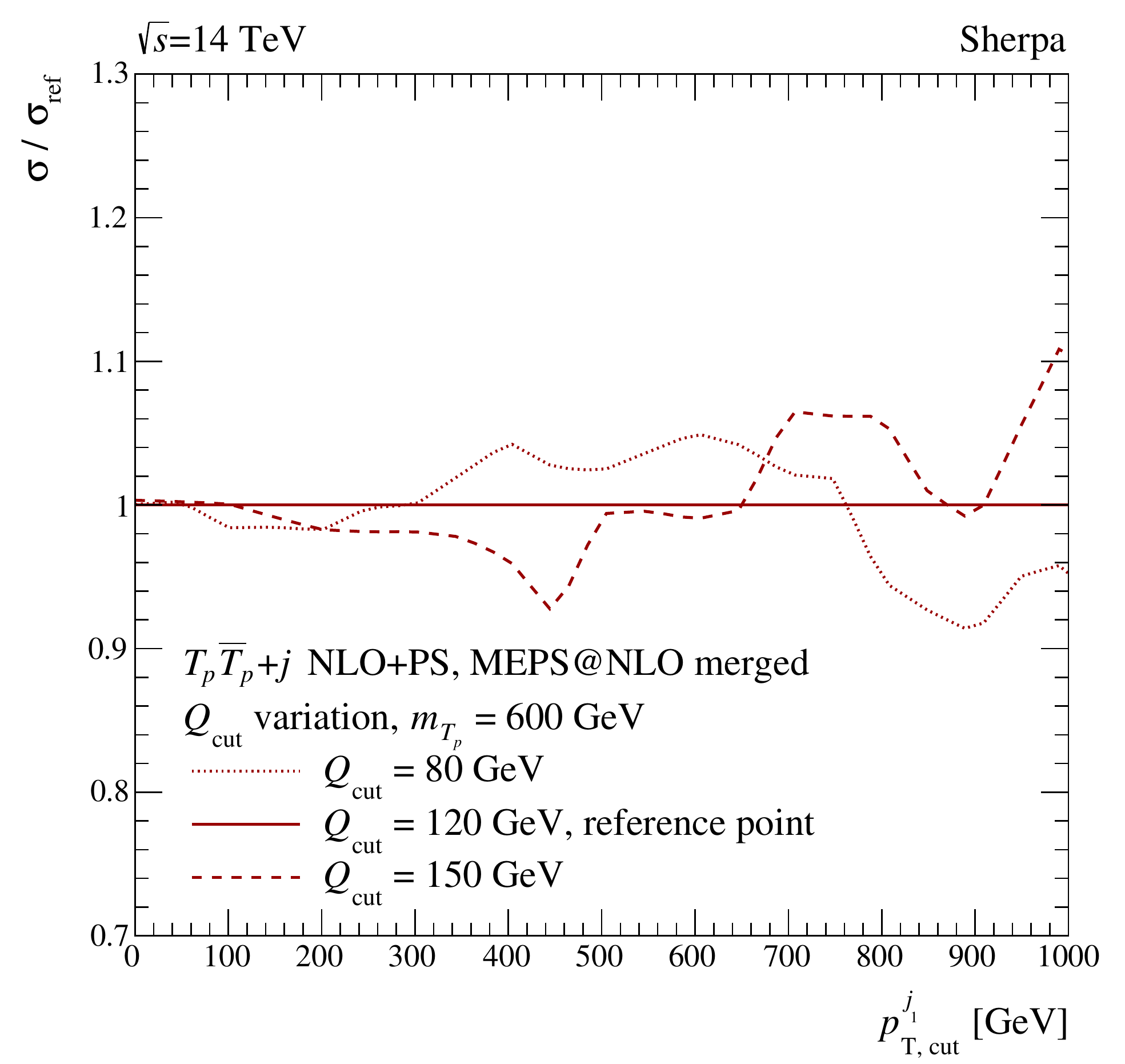}\caption{}\label{fig:meps-PDF-QCut}
  \end{subfigure}
  \caption{The normalized (a) PDF uncertainty and (b) MEPS@NLO jet merging scale $\Qcut$ dependence
  for the $\pptptpj$ cross section as a function of $\ptjcut$, as derived from the  $1\jet$ MEPS@NLO calculation.
  }
  \label{fig:meps-PDF}
\end{figure*}

Having established that multijet-merged simulations with NLO matrix elements feature
moderate-to-small factorization, renormalization, and shower scale dependencies,
we lastly investigate the uncertainty associated with our PDF input and the MEPS@NLO merging scale.
In \figref{fig:meps-PDF-PDF}, we show again the normalized $\pptptpj$ cross section,
as derived from the $\tptp$ process with MEPS@NLO merging to one additional jet, and the associated $1\sigma$ PDF variation band.
The uncertainty is derived from PDF replicas, following the procedure of Ref.~\cite{Buckley:2014ana}.
For the range of $\ptjcut$ investigated, we find that the band is stable, symmetric, 
and spans roughly $\pm6\%$ at low $\ptjcut$ to $\pm10\%$ at high $\ptjcut$.
At 27 TeV, we largely investigate a comparable range of Bjorken-$x$ as we do at 13 TeV, 
and hence expect the PDF uncertainty to remain the same due to the stability of DGLAP evolution.
In \figref{fig:meps-PDF-QCut}, we show the uncertainty associated with the MEPS@NLO jet merging scale, $\Qcut$.
Over a considerable range of $\ptjcut$, we observe variations below 
$\pm5\%$ for $\ptjcut<400\GeV$ and below $10\%$ for $\ptjcut<1\TeV$, and is comparable to our MC uncertainty.


\section{Outlook}
\label{sec:outlook}
In \sectionref{sec:ident}, we discussed the dependence of the $\ppQQbarj$ cross section 
on the mass, spin, and color representation of $\Q$. 
A single signal cross section measurement does not, of course, constrain the nature of $\Q$ 
uniquely because different spin and color hypotheses can lead to identical cross sections if the mass is chosen/tuned accordingly. 
For example: in the representative case above, 
a stop of mass \SI{400}{\giga\electronvolt}, 
a fermionic top partner of mass \SI{600}{\giga\electronvolt}, 
and a gluino of mass \SI{800}{\giga\electronvolt} have practically the same $\ppQQbarj$ cross section 
for $\ptjcut \approx \SI{600}{\giga\electronvolt}$ at $\sqrt{s}=\SI{14}{\tera\electronvolt}$. 
The degeneracy can be resolved, however, through additional cross section measurements with larger $\ptjcut$ and/or at a higher center-of-mass energy. 
Due to the different masses of the representative cases, the cross sections scale differently with increasing jet transverse momentum and the energy.
This leads to a break in the degeneracy, thereby enabling one to discriminate against the various $\Q$ hypotheses.

In the left panel of \figref{fig:ptj-dependence}, we show that all three cases could be distinguished through a secondary measurement
of the $\ppQQbarj$ cross section at $ \ptjcut=\SI{1}{\tera\electronvolt}$ 
were our predictions were precise at the \SI{10}{\percent} level.
The nominal difference between the gluino and fermionic top partner case is roughly that magnitude, 
while for the case of scalar top, the difference could reach 50\%.
In light of the uncertainties estimated in \sectionref{sec:theoUncertainty}, it is unrealistic
to claim that one can distinguish a gluino from top partner in this manner, 
with renormalization and factorization scale uncertainties alone reaching about \SI{20}{\percent}. 
Additionally one can also make use of the higher energy collisions, 
viz. the potential 27 TeV upgrade of LHC, to distinguish these three particles. 
As shown in \figref{fig:ptj-dependence}(b),
the degeneracy in cross sections observed for $\ptjcut$ = 500 GeV at $\sqrt s $ = 14 TeV is lifted at $\sqrt s $ = 27 TeV.
Hence, additional information for discriminating against competing $\Q$ hypotheses is made available by combining data at 14 TeV and 27 TeV.

Alternatively, instead of considering cross sections themselves, one could also consider ratios of the form
\begin{align}
  \frac{\sigma( \ptj > \ptjcut)}{\sigma( \ptj
  > \ptjcut{}^\prime)}\,.
  \label{eq:xs-ratio}
\end{align}
As discussed in \sectionref{sec:ident}, ratios of cross sections measured at
$\sqrt{s}=\SI{14}{\tera\electronvolt}$ and $\sqrt{s}=\SI{27}{\tera\electronvolt}$ 
can also effectively discriminate against various $\Q$ hypotheses
if theoretical uncertainties on these ratios are smaller than about \SI{30}{\percent}.

\begin{figure*}
\begin{center}
  \begin{subfigure}{.49\textwidth}
  \includegraphics[width=\textwidth]{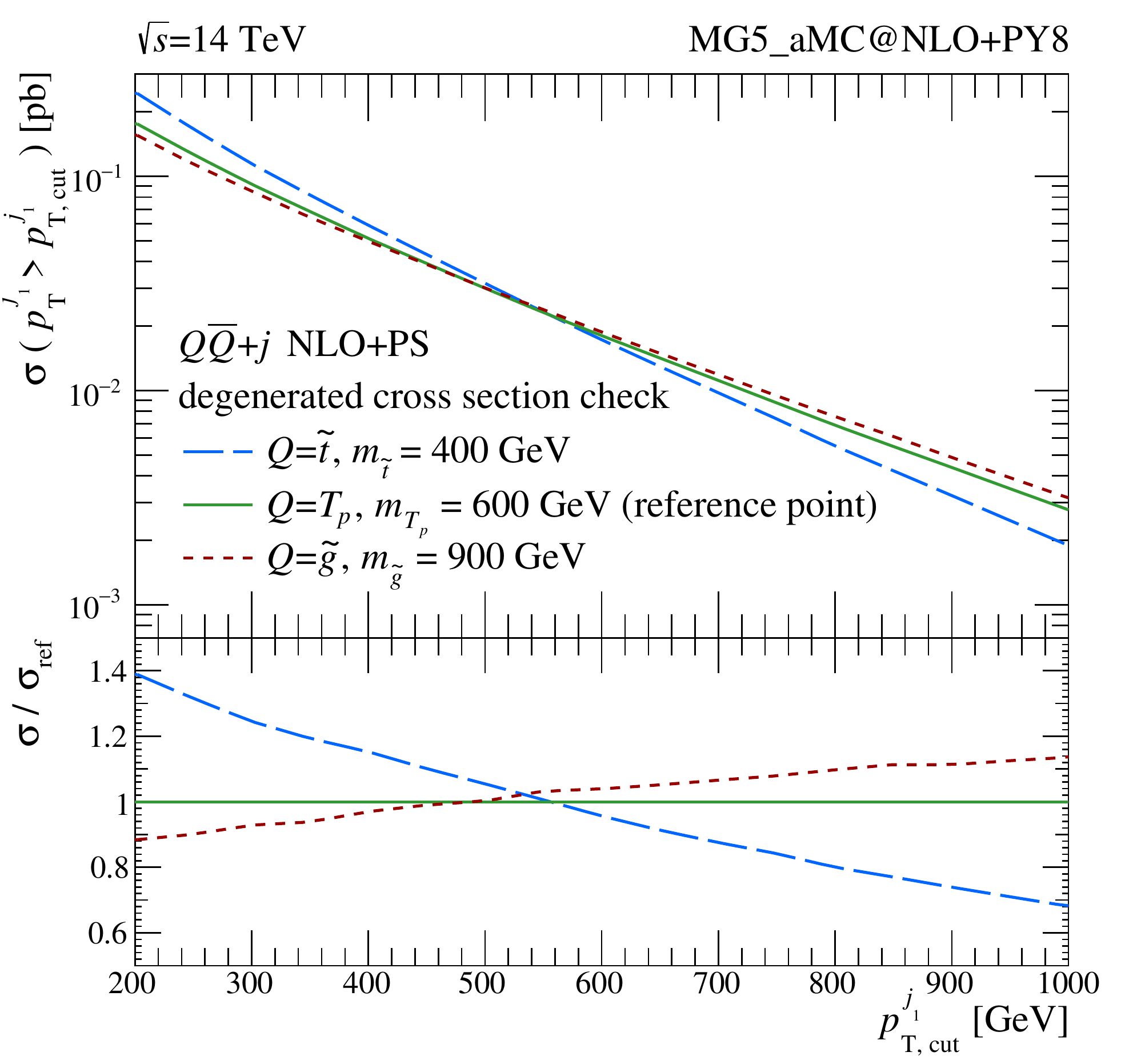} \caption{}
  \end{subfigure}
  \hfill
  \begin{subfigure}{.49\textwidth}
  \includegraphics[width=\textwidth]{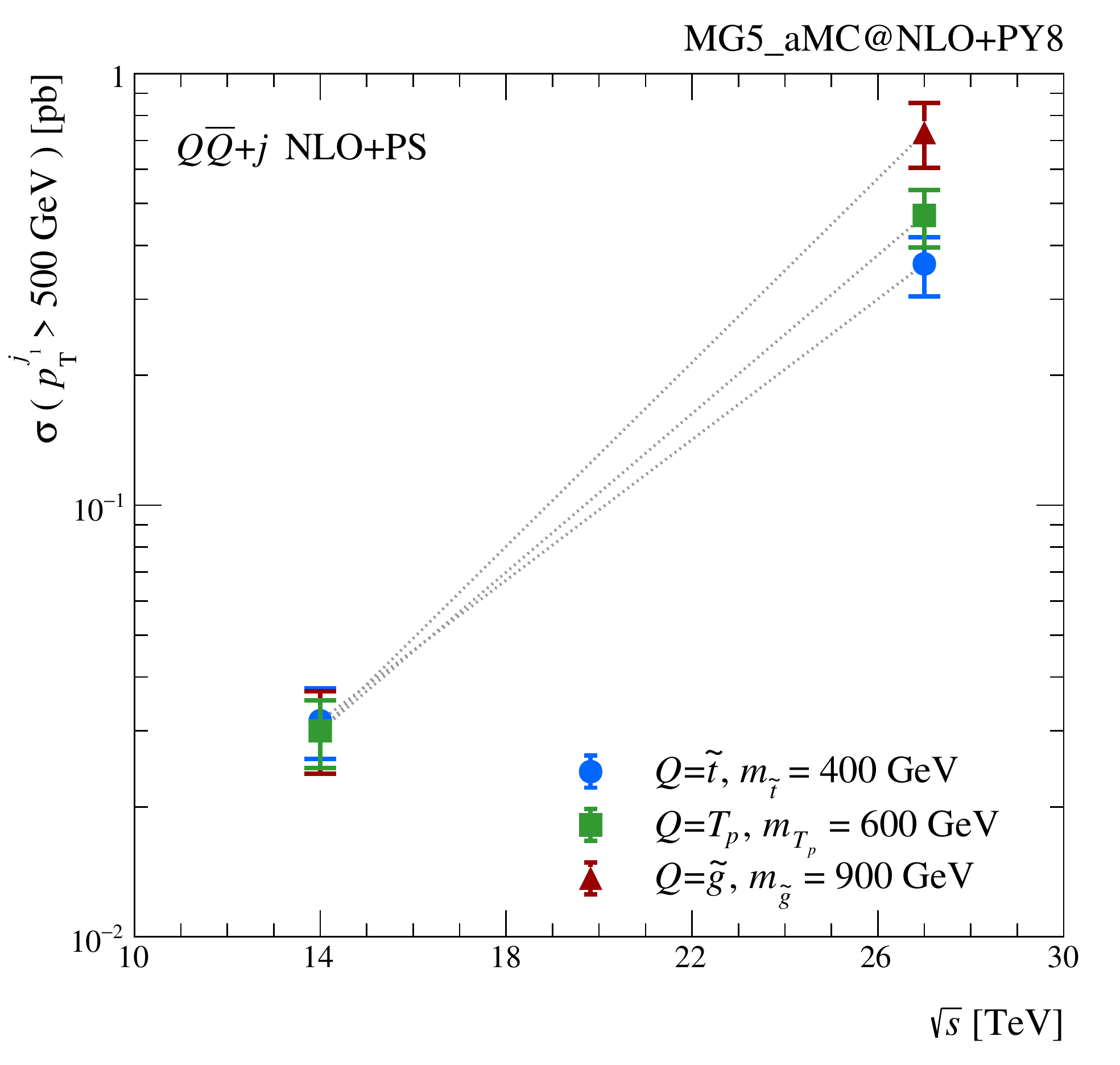} \caption{}
   \end{subfigure}
     \caption{
(a) Upper: The $\ppQQbarj$ cross section at $\sqrt{s}=14\TeV$ 
as a function of jet $\pt$ selection criterion $(\ptjcut)$, for representative $(\Q,m_{\Q})$ combinations.
Lower: The same but normalized to the $(\Q,m_{\Q})=(\tp,600\GeV)$ curve.
(b) The $\ppQQbarj$ cross section at $\sqrt{s} =14$ and 27 TeV, with $\ptjcut$ = 500 GeV. 
The error bar reflects the renormalization and factorization scale variation.
     }
\label{fig:ptj-dependence}
\end{center}
\end{figure*} 

In ratios and double ratios, such as those discussed in previous sections, the normalization component of uncertainties cancel. 
As shown in figure \ref{fig:shvsmgL}, for example, variations of the factorization and renormalization scales 
affect mainly the overall normalization, not the shape/$\pt^\jet$ dependence.
Therefore, uncertainties estimated in this way are expected to drop out in \eqref{eq:xs-ratio}.
To verify this, we calculate the scale uncertainty for the ratio of the cross section to the nominal scale choice:
\begin{eqnarray}
\label{eq:xs-ratio2}
& & {\mathcal F}(\ptjcut, p^{\star}, \mu_R, \mu_F)= \\
& & \frac{\sigma(\ptj > \ptjcut, Q_R=\mu_R Q_0, Q_F=\mu_F Q_0)/\sigma(\ptj  > p^{\star}, Q_R=\mu_R Q_0, Q_F=\mu_F Q_0)}
{
\sigma(\ptj > \ptjcut, Q_R=Q_0, Q_F= Q_0)/ \sigma( \ptj  > p^{\star}, Q_R=Q_0, Q_F= Q_0)} \nonumber
\,.
\end{eqnarray}
For the normalizing cross section, we consider the  $m_{\tp}=600$~GeV and vary the factorization and renormalization scales in the same manner
and choose $p_{\star}=300$ GeV. 
We observe that the effect of varying $\mu_{R,F}$ between 1/2 and 2 with $Q_0=H_T/2$ is marginal (only 2\%)  at 1~TeV. 
The PDF uncertainty, however, impacts the cross section more, though the effect remains no more than 10\% at 1 TeV. 

\begin{figure*}[!t]
  \centering
  \begin{subfigure}{.49\textwidth}
  \includegraphics[width=\textwidth]{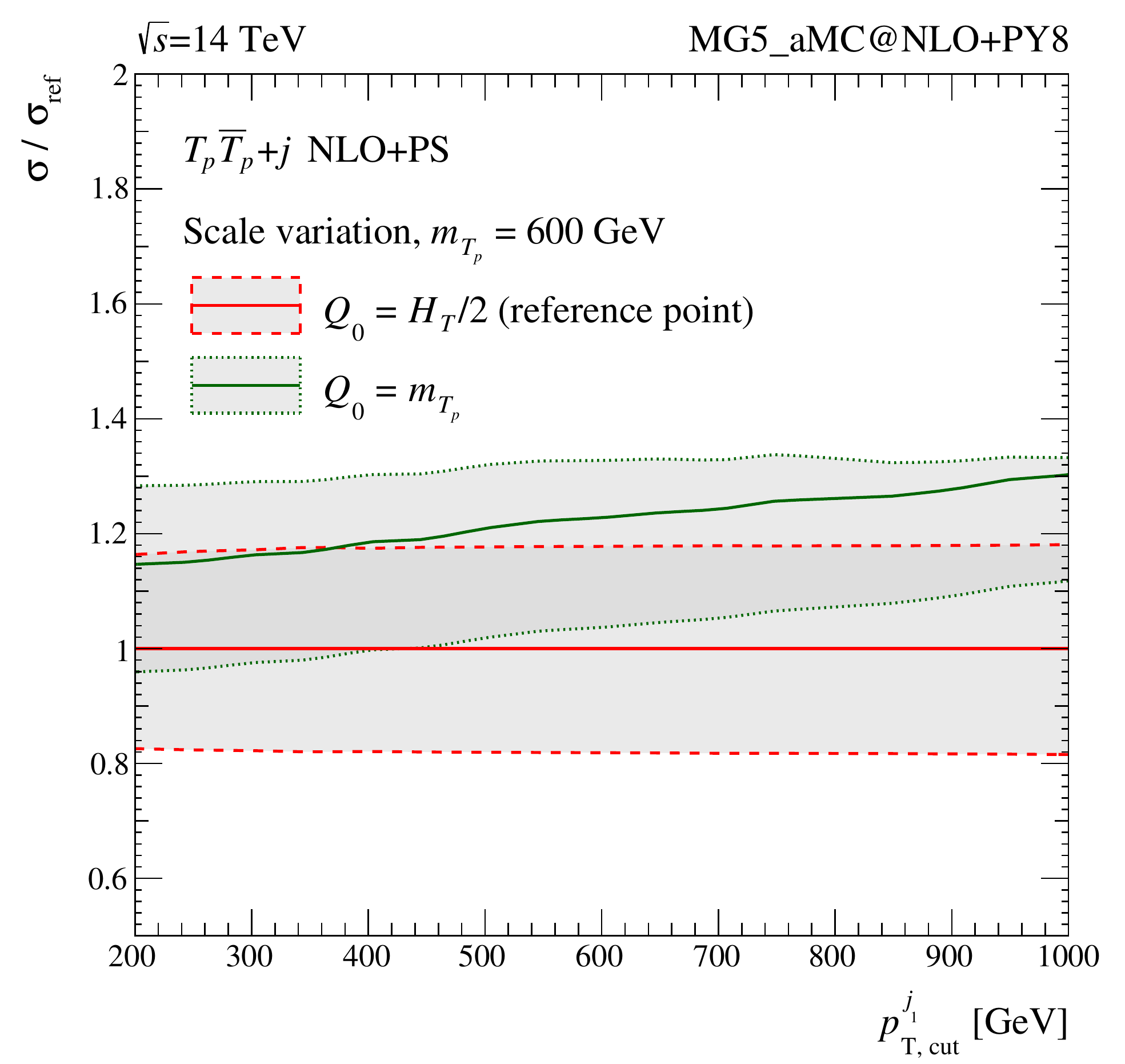}\caption{}
  \end{subfigure}
  \hfill
  \begin{subfigure}{.49\textwidth}
  \includegraphics[width=\textwidth]{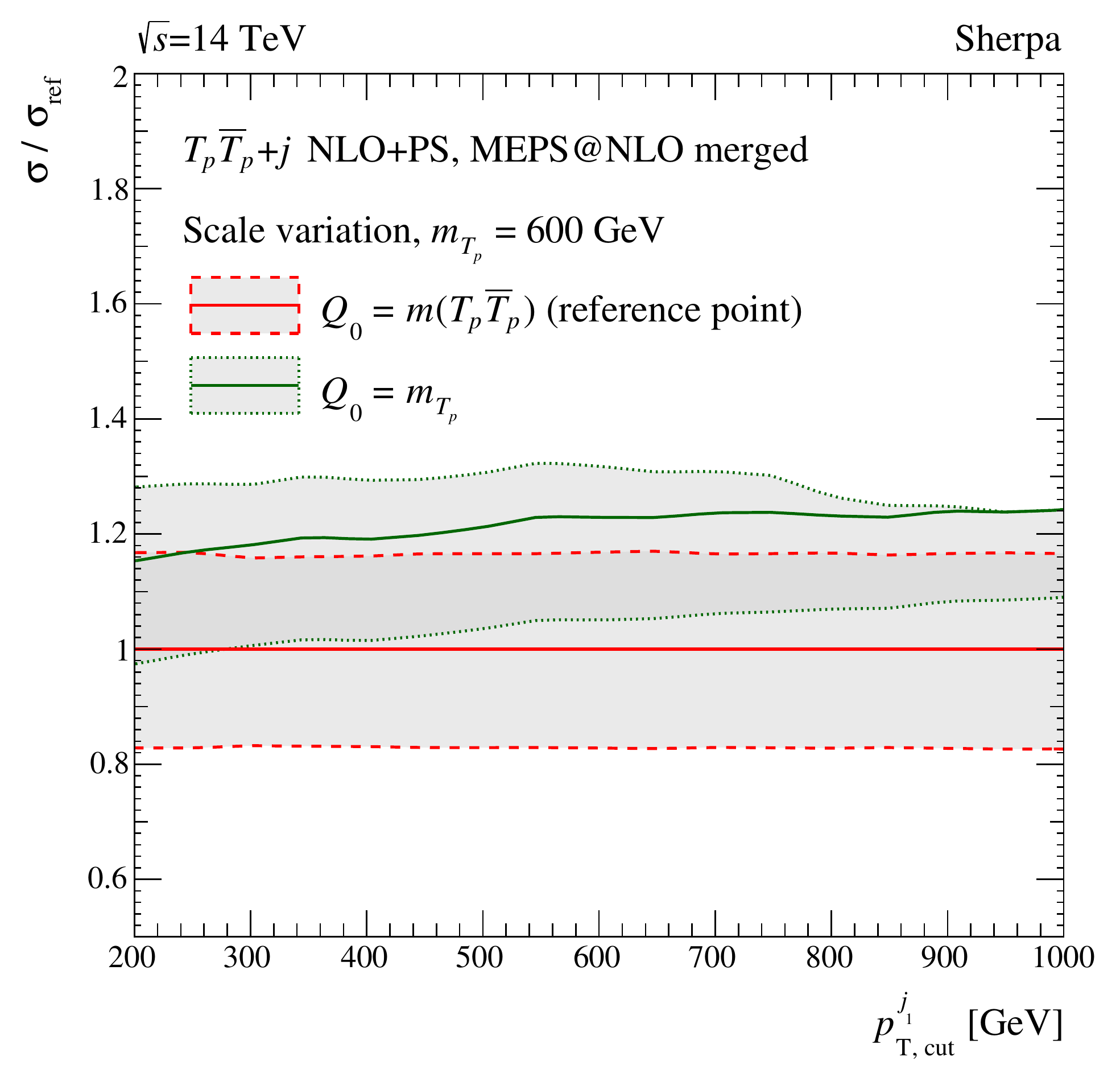}\caption{}
  \end{subfigure}
\caption{The effect of choosing an alternative scale choice $Q_0 = m_{\tp}$ for $\tptpj$ process using \mgpy~(left) and \sherpa~(right). 
The envelopes denote scale uncertainty bands. The distributions are normalized to nominal scale $Q_0 = H_T/2$. }
\label{fig:slope}
\end{figure*}

In Eq.~(\ref{eq:xs-ratio2}), we assume the scale choices are correlated for both the numerators and denominators. 
Fairly, one might argue that in  order to estimate uncertainties on cross section ratios through the shape of the $\ptjcut$ dependence, 
one should consider different functional forms of $Q_R$ and $Q_F$ as well. 
In \figref{fig:slope} we show that the effect of choosing a non-dynamical nominal scale can in fact induce significant slope variations. 
Unsurprisingly, deviations are observed for both \mgpy~ and \sherpa~events. 
The ratio of the cross section at 1~TeV and 300~GeV for $Q_0=m_{\tp}$ is about 15\% higher than $Q_0=H_T/2$. 
The alternative scale choice $Q_R = Q_F = m_{\tp}$ is, however, also quite extreme in the sense that it is completely static~\cite{Berger:2009ep}. 
For example: the upper envelope for the scale variation (i.e., $\mu_{R} = \mu_F = 0.5$) is calculated with $Q_R =Q_F = 300$~GeV, 
but the event with $\ptjcut=1$~TeV implies $\sqrt{s}> 2$ TeV. 
Therefore, the ratio $\sqrt{s}/Q_R$ is a factor of 7, which leads to large non-asymmetric effects as seen in the right panel of \figref{fig:slope}.


We conclude that cross section ratios, in general, will provide a more robust way for discrimination but also that the assessment of theoretical uncertainties on such ratios is a highly nontrivial issue. 
It is natural to expect a majority of the scale uncertainty will cancel in the cross section ratio. 
However, in general, an NNLO calculation can induce correction terms that scale with jet $p_T$, and result in an effect may not be small. We have seen in \figref{fig:shvsmgL} that different reference scale choices affect the cross section ratio at the level of 10\%. The assessment of theoretical uncertainty on such ratio can only be estimated reliably by incorporating  higher order calculations.  
A more quantitative study of this problem is, however, beyond the scope of this paper.

As a final comment, we note that more precise predictions for cross sections can in any case help to improve the discriminating power 
of the measurements proposed here. 
Presently, NNLO in QCD predictions for the relevant processes (which include an additional jet in the final state) are not available.
For the inclusive process $\pptptp$, however, NNLO predictions can be obtained 
from \hathor~\cite{Aliev:2010zk} and \texttt{top++}~\cite{Czakon:2013goa}. 
In \figref{fig:xsecCompare} we show that the uncertainties on the total cross section are 
approximately halved when going from NLO to NNLO in $\alpha_s$. 
Note that the uncertainty for the $\pptptpjj$ LO matched cross-section is around 25-30\%, while the uncertainty reduces to 10\% at NLO. 
Moreover, the uncertainty in NLO cross-section estimated by three independent sources also show excellent agreement. 
(Note that for \sherpa, we use the scale scheme $Q_F = Q_R = m(\tptp)/2$ 
to approximate the \mgamc~scale scheme of $Q_F = Q_R = H_T /2$.)
The NNLO uncertainty on the total cross section is only around \SI{5}{\percent}.

\begin{figure}[!t]
\begin{center}
\includegraphics[width=0.7\textwidth]{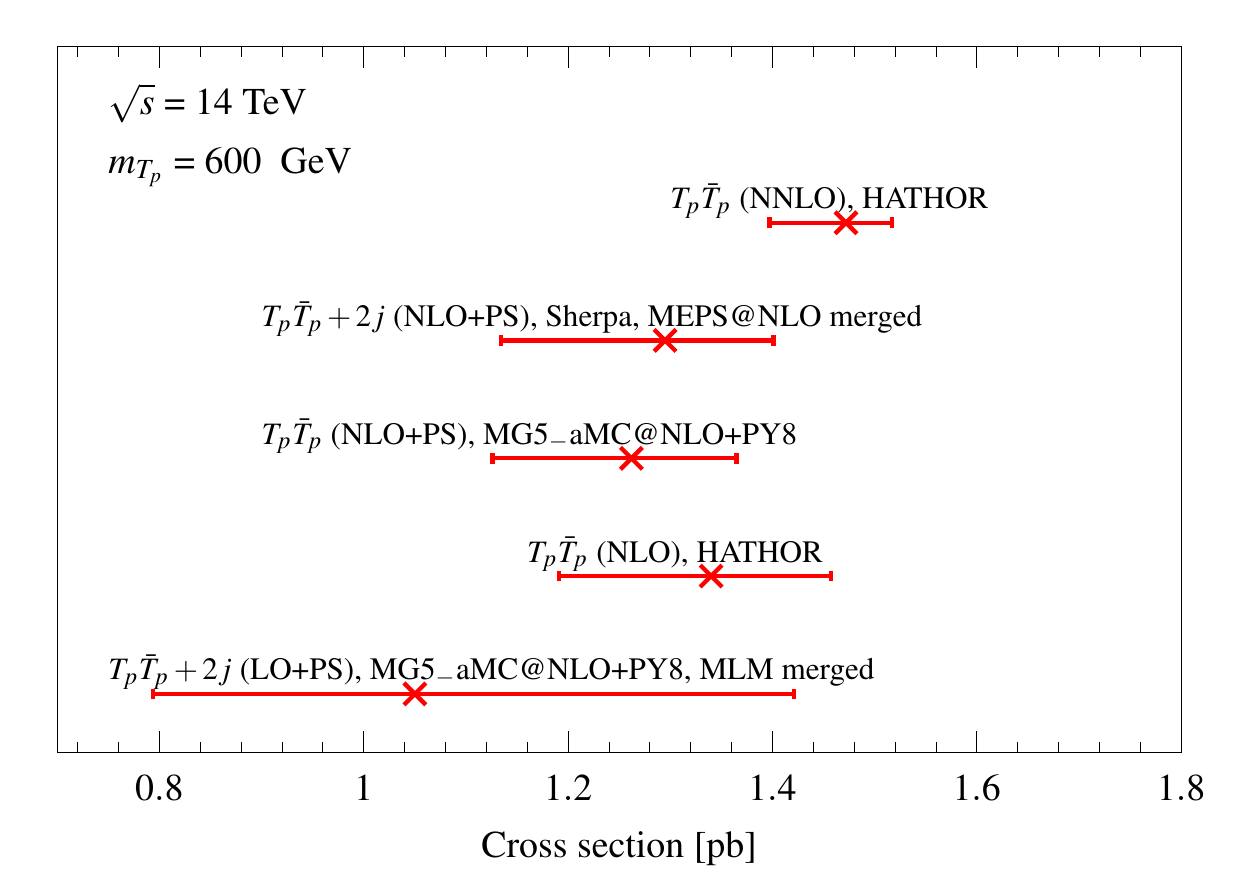}
\caption{Cross section for the $\pptptp$ process with additional 
light QCD partons at the ME level at leading, next-to-leading and next-to-next-to-leading orders 
in $\alpha_s$ provided by the available MC tools. 
}
\label{fig:xsecCompare}
\end{center}
\end{figure}

Uncertainties on monojet signal cross section are expected to be larger than the uncertainty on the inclusive $\pptptp$ process. 
In \figref{fig:test}, we show that renormalization and factorization scale uncertainties are around \SI{40}{\percent} at LO and \SI{20}{\percent} at NLO, 
which is twice as high as for the inclusive $\pptptp$ cross section. 
This is to be expected because the lowest order term of the total cross section is proportional to $\alpha_s^2$ in the inclusive case, 
while the monojet signal cross section is proportional to $\alpha_s^3$.
Assuming that a comparable relative gain can be achieved for $\pptptpj$ when going to NNLO, 
it might be possible to get to a theoretical precision of about \SI{10}{\percent} when NNLO calculations become available. 
Considering the large momentum transfers involved, 
EW corrections might have to be added as well in order to capture the effect of (real and virtual) EW Sudakov 
logarithms~\cite{Denner:1991kt,Bauer:2016kkv,Biedermann:2017yoi,Frederix:2018nkq}.

\section{Summary}\label{sec:summary}
We have studied the monojet collider signature arising from the process $\ppQQbarj$, 
where $\Q$ is a heavy colored particle that decays into an invisible particle $\dm$ with mass $m_\dm$ close to the mass of $\Q~(m_{\Q})$.  
In the limit where $(m_{\Q}-m_\dm)/m_{\Q}$ is small, the full process signature is monojet-like, 
and the discovery prospects for $\Q$ and $\dm$ become much more challenging, even at the HL-LHC or proposed HE-LHC, due to large SM backgrounds.

In this context, we have investigated the feasibility of extracting the properties of $\Q$ were a monojet signature discovered,
focusing on the observation that the monojet signal cross section $\sigma(\ppQQbarj)$ 
is sensitive to the mass, spin, color representation of the particle.
Due to an interplay of these quantities, one cannot readily and uniquely determine the nature of the particle $\Q$ 
from a single cross section measurement alone without assuming the mass of the particle.

We have studied several processes calculated at next-to-leading order in QCD with parton 
shower matching and multijet merging using the state-of-the-art Monte Carlo suites \mgpy~and~\sherpa. 
It has been pointed out that the dependence on $\pt^\jet$ in the $\ppQQbarj$ cross section, up to overall normalization,
depends {considerably} on the mass of $\Q$ but exhibits a much milder dependence on its color representation.
Moreover, we find that there is {little-to-no} spin dependence in the {shape of the cross section curves}.　 
The dependencies of the overall normalization and $\pt^\jet$ on the underlying parameters can, in principle,
allow efficient discrimination of the nature of  $\Q$, and hence the underlying physics behind the monojet signature.

Were the process discovered, there are many obstacles to extracting the cross section of the monojet signature and its $\pt$ dependence.  
First, the signal is overwhelmed by the background.
Therefore, large statistics and a precise understanding of the background is required to observe the signature
at the HL-LHC for the parameter region that has not been excluded by current LHC data.   
Increasing the collider energy improves sensitivity significantly, and represents a promising possibility.
In addition, theoretical predictions of the overall cross section normalization need to be sufficiently accurate
in order to differentiate against various candidate scenarios.
For the relevant model parameter regions investigated, we observe that the overall 
cross sections need to be  predicted within an uncertainty of $\Delta \sigma/\sigma <30$\%, 
while $\pt^\jet$ dependencies should be known within {$\delta \sigma / \delta \pt^\jet < 10\%/100\GeV$}. 
Finally, we find that the achievable precision for such quantities is comparable to the required precision.


\begin{acknowledgements}

The authors would like to thank Bryan Webber, Frank Krauss, Olivier Mattelaer, Rikkert Fredrix, Fifo Tamarit, Paolo Torrielli, Marco Zaro 
for many useful comments and discussions. 
This work supported by the Grant-in-Aid for Scientific Research on Scientific Research B 
(No.16H03991, 17H02878 [MMN]) and Innovative Areas (16H06492 [MMN]), and by World Premier 
International Research Center Initiative (WPI Initiative), MEXT, Japan. 
The work of SHL was supported in part by MEXT KAKENHI Grant Number JP16K21730.
The work of RR was funded in part by the UK Science and Technology Facilities Council (STFC),
the European Union's Horizon 2020 research and innovation programme under the Marie Sklodowska-Curie grant agreements
No 690575 (InvisiblesPlus RISE) and No 674896~(InvisiblesPlus RISE).
SK's work is supported by the U.S. Department of Energy under contract number DE-AC02-76SF00515.
RR acknowledges the generous hospitality of KEK and IPMU during the starting stages of this work.
\end{acknowledgements}

\bibliographystyle{spphys}
\bibliography{reference}

\end{document}  
